\documentclass[aps,pra,reprint,showpacs,superscriptaddress,floatfix]{revtex4-2}
\usepackage{placeins}
\usepackage{amsmath}
\usepackage{amssymb}
\usepackage{mathtools}
\usepackage[utf8x]{inputenc}
\usepackage{natbib}
\usepackage{graphicx}
\usepackage{epstopdf}
\usepackage{pifont}
\usepackage{ulem}

\usepackage{hyperref}
\usepackage{cleveref}
\usepackage{physics}
\usepackage{soul}
\usepackage[export]{adjustbox}
\usepackage{orcidlink}
\graphicspath{{pdf/}}
\hypersetup{colorlinks=true,linkcolor=blue,citecolor=blue,filecolor=blue,urlcolor=blue,breaklinks=true}
\RequirePackage{color}

\begin{document}

\date{\today}

\title{Helically twisted spacetime: study of geometric and wave optics, and physical analysis}

\author{Edilberto O. Silva\orcidlink{0000-0002-0297-5747}}
\email[Edilberto O. Silva - ]{edilberto.silva@ufma.br}
\affiliation{Departamento de F\'{\i}sica, Universidade Federal do Maranh\~{a}o, 65085-580 S\~{a}o Lu\'{\i}s, Maranh\~{a}o, Brazil}

\author{Frankbelson dos S. Azevedo\orcidlink{0000-0002-4009-0720}}
\email[Frankbelson dos S. Azevedo - ]{frfisico@gmail.com}
\altaffiliation{On leaving the current affiliation and currently unaffiliated.}
\affiliation{Departamento de F\'{\i}sica, Universidade Federal do Maranh\~{a}o, 65085-580 S\~{a}o Lu\'{\i}s, Maranh\~{a}o, Brazil}

\author{Faizuddin Ahmed\orcidlink{0000-0003-2196-9622}}
\email[Faizuddin Ahmed - ]{faizuddinahmed15@gmail.com}
\affiliation{Department of Physics, Royal Global University, Guwahati, 781035, Assam, India}

\begin{abstract}
We analyse a stationary, cylindrically symmetric spacetime endowed with an intrinsic helical twist, $ds^{2} = -dt^{2} + dr^{2} + r^{2} d\phi^{2} + (dz + \omega\, r\,d\phi)^{2}$. Solving the Einstein equations exactly yields an anisotropic energy--momentum tensor whose density is negative and decays as $r^{-2}$, thus violating the weak energy condition near the axis.  
Three notable features emerge:  
(i) axis‐centred negative energy;  
(ii) unequal transverse stresses;  
(iii) a torsional momentum flux $T_{\phi z}\propto \omega^{3}/r$.  
We identify stable photon orbits and deflection angle, fully helical geodesics, and torsion-controlled wave optics modes, suggesting laboratory analogues in twisted liquid‐crystal and photonic systems. 
The coupling between the torsion parameter $\omega$ and other physical parameters leads to significant effects, altering the motion along the positive or negative $z$-axis.
These results make the twisted helical metric a useful test bed for studying the interplay of curvature, torsion, and matter in both gravitational and condensed-matter contexts.
\end{abstract}

\maketitle

\section{Introduction \label{sec:intro}}

{\small

Einstein’s General Theory of Relativity (GR) stands as one of the foundational pillars of modern physics, providing the most accurate and comprehensive description of gravity to date. It fundamentally reinterprets gravity not as a force, but as a manifestation of the curvature of spacetime caused by the presence of mass and energy. Over the past century, GR has withstood a wide range of experimental and observational tests with remarkable success. These include the precise explanation of Mercury’s perihelion precession~\cite{FA1}, the phenomenon of gravitational redshift~\cite{FA2}, the behavior of binary pulsars composed of solar-mass neutron stars~\cite{FA3}, and the detection of gravitational waves from stellar-mass black hole mergers~\cite{FA4}. A recent milestone supporting GR is the first direct observation of Schwarzschild precession in the orbit of the star S2 around the supermassive black hole candidate at the center of our galaxy, SgrA*~\cite{FA5}. Additionally, the Event Horizon Telescope’s groundbreaking imaging of the photon sphere, the bright emission ring, and the shadow of the supermassive black hole in M87~\cite{FA6} has further captivated the scientific community, offering unprecedented visual evidence consistent with the predictions of general relativity. Numerous exact solutions of the field equations were reported in general relativity, which possesses many interesting geometrical and physical properties. For detailed discussion of the exact solutions, readers are advised to see these references \cite{HS,JBG}.

Incorporating twisting effects in spacetime models with helical structures is a significant area of investigation. For example, geometric effects on the electronic structure of helical curved geometries, such as nanotubes and graphene, have been explored through geometrically induced potentials in the effective Schrödinger equation \cite{de2021geometric}.  The combined effects of cosmic string spacetime and helical geometry (dislocation) reveal a rich particle behavior structure, influencing particle motion and wave propagation \cite{dos2024dynamics}. Earlier studies by Klein \cite{PhysRevD.70.124026} investigated binary black hole spacetimes with a helical Killing vector, revealing oscillatory behavior in metric functions and the absence of a smooth null infinity. Helical structures, with torsional degrees of freedom, crucial in both gravity and condensed matter, can support topological ``monopole'' configurations with integer-valued properties \cite{Randono2010TorsionalMA}. Dislocations in crystals, which carry torsion, offer opportunities for experimental exploration in gravitational contexts, since the search for torsion in gravity ventures into largely uncharted territory. This opens a promising new avenue for studying topological defects with helical structures in soft matter physics \cite{Randono2010TorsionalMA,RichardTHammond_2002}.

The interplay between spacetime geometry and condensed matter physics is central to GR, particularly through topological defects and geometric approaches \cite{moraes2000condensed,doi:10.1126/science.263.5149.943}. The field of topological states in condensed matter has experienced rapid growth, encompassing topological insulators, the quantum anomalous Hall effect, chiral and helical topological superconductors, and Weyl semimetals \cite{wang2017topological}. The intersection of topology/geometry and condensed matter has bridged disciplines, providing rigorous mathematical frameworks while remaining accessible to physicists \cite{bhattacharjee2017topology}.

Therefore, in this work, we analyse an exact solution to Einstein's field equations characterized by a helical spatial structure. This geometry belongs to the class of stationary, cylindrically symmetric spacetimes but includes a non-trivial coupling between the axial and angular directions, introduced through a torsional parameter $\omega$. The metric under consideration is fully four-dimensional, with a line element given by:
\begin{equation}
ds^2 = -dt^2 + dr^2 + r^2\, d\phi^2 + (dz + \omega\, r\, d\phi)^2, \label{metric}
\end{equation}
where $\omega$ is a real dimensionless parameter. The term $(dz + \omega r d\phi)^2$ captures the geometric twist of the spacetime, inducing an effective torsion without relying on Einstein-Cartan theory. The metric describes a spacetime with a built-in helical twist, creating torsion-like effects purely from its geometry \cite{KATANAEV19921,RolandAPuntigam_1997}.

It is worth emphasizing that the torsional parameter $\omega$ may take either positive or negative values. Physically, $\omega$ controls the handedness of the helical structure embedded in the spatial geometry: a positive $\omega$ corresponds to a right-handed helicoid, while a negative $\omega$ corresponds to a left-handed one. Moreover, the presence of both signs is consistent with physical realizations of helical dislocations in condensed matter analogues and twisted spacetime models \cite{moraes2000condensed,dos2024dynamics}. This distinction may become particularly relevant when analyzing geodesic motion, as the coupling between angular and linear momentum (along the $z$-direction) and torsion may lead to modifications that depend on the sign of $\omega$. Ultimately, it is not merely the sign of $\omega$ that determines whether a particle follows a left- or right-handed helicoidal trajectory, but how $\omega$ couples with the angular and linear momenta along the $z$-axis. In fact, this matter will be fully analyzed later in the paper when we study the dynamics of geodesics and wave optics. Namely, throughout this work, we treat $\omega$ as a free parameter and explore its magnitude. Changes in its sign will be analyzed in conjunction with the physical parameters that together define the direction of motion along the positive or negative $z$-axis.

The paper is organized as follows.  In Sec.~\ref{geometry} we introduce the metric, derive the full energy--momentum tensor and discuss the relevant energy conditions.
Section~\ref{sec:geodesics} is devoted to the three--dimensional geodesic structure, while Sec.~\ref{sec:hamiltonian} reformulates the problem in Hamiltonian form, analysing orbit stability, special trajectories, and the
effective potential. In Sec.~\ref{deflection} we compute the deflection angle for photon rays,
and in Sec.~\ref{sec:waveoptics} we investigate wave–optics phenomena in the helically twisted background. Finally, Sec.~\ref{sec:conclusion} summarises our main results and outlines
prospects for future works.

\section{Geometric Foundations and Energy Conditions}\label{geometry}

We now present the full geometric structure of the spacetime defined by the line element \eqref{metric}.
We begin by solving the Einstein field equations,
\begin{equation}
G_{\mu\nu} = R_{\mu\nu} - \frac{1}{2} R g_{\mu\nu} = 8\pi T_{\mu\nu},
\end{equation}
to determine the energy-momentum tensor $T_{\mu\nu}$ that sources this geometry. Our analysis proceeds by computing the Christoffel symbols, curvature tensors, and finally the Einstein tensor, from which we extract the physical content implied by this helical background.

The metric components and their inverse are
\begin{align}
g_{\mu\nu} &= \begin{pmatrix}
-1 & 0 & 0 & 0 \\
0 & 1 & 0 & 0 \\
0 & 0 & r^2(1 + \omega^2) & \omega\, r \\
0 & 0 & \omega\, r & 1
\end{pmatrix}, \notag \\
g^{\mu\nu} &= \begin{pmatrix}
-1 & 0 & 0 & 0 \\
0 & 1 & 0 & 0 \\
0 & 0 & \frac{1}{r^2} & -\frac{\omega}{r} \\
0 & 0 & -\frac{\omega}{r} & 1 + \omega^2
\end{pmatrix}.\label{metric-tensor}
\end{align}
Using this metric tensor, we compute the non-vanishing Christoffel symbols of the second kind in the 3+1 spacetime as follows:
\begin{align}
\Gamma^1_{22} &= -r(1 + \omega^2), & \Gamma^1_{23} &= \Gamma^1_{32} = -\frac{\omega}{2}, \nonumber\\
\Gamma^2_{12} &= \Gamma^2_{21} = \frac{2 + \omega^2}{2r}, & \Gamma^2_{13} &= \Gamma^2_{31} = \frac{\omega}{2r^2}, \nonumber\\
\Gamma^3_{12} &= \Gamma^3_{21} = -\frac{\omega(1 + \omega^2)}{2}, & \Gamma^3_{13} &= \Gamma^3_{31} = -\frac{\omega^2}{2r}.
\end{align}
All remaining components vanish. The symmetry $\Gamma^\lambda_{\mu\nu} = \Gamma^\lambda_{\nu\mu}$ holds, and dimensional consistency confirms that all symbols have units of inverse length. This structure reduces to flat spacetime in cylindrical coordinates when $\omega \to 0$. The radial dependence $\sim 1/r$ and $\sim 1/r^2$ reflects coordinate singularities and curvature concentrated near the axis.

The Riemann tensor follows from
\begin{equation}
R^\rho_{\sigma\mu\nu} = \partial_\mu \Gamma^\rho_{\nu\sigma} - \partial_\nu \Gamma^\rho_{\mu\sigma} + \Gamma^\rho_{\mu\lambda} \Gamma^\lambda_{\nu\sigma} - \Gamma^\rho_{\nu\lambda} \Gamma^\lambda_{\mu\sigma},
\end{equation}
and yields the Ricci tensor
\begin{equation}
R_{\mu\nu} = \begin{pmatrix}
0 & 0 & 0 & 0 \\
0 & -\dfrac{\omega^2}{2r^2} & 0 & 0 \\
0 & 0 & \dfrac{\omega^2(1 + \omega^2)}{2} & \dfrac{\omega(1 + \omega^2)}{2r} \\
0 & 0 & \dfrac{\omega(1 + \omega^2)}{2r} & \dfrac{\omega^2}{2r^2}
\end{pmatrix},
\end{equation}
and scalar curvature
\begin{equation}
R = -\frac{\omega^2}{2r^2}.
\end{equation}

The Einstein tensor then follows as
\begin{equation}
G_{\mu\nu} = \begin{pmatrix}
-\dfrac{\omega^2}{4r^2} & 0 & 0 & 0 \\
0 & -\dfrac{\omega^2}{4r^2} & 0 & 0 \\
0 & 0 & \dfrac{3\omega^2(1 + \omega^2)}{4} & \dfrac{\omega(3\omega^2 + 2)}{4r} \\
0 & 0 & \dfrac{\omega(3\omega^2 + 2)}{4r} & \dfrac{3\omega^2}{4r^2}
\end{pmatrix},
\end{equation}
which leads directly to interpreting the matter content through the energy-momentum tensor $T_{\mu\nu} = G_{\mu\nu} / 8\pi$.

The energy-momentum tensor implied by this geometry is anisotropic and exhibits both radial dependence and off-diagonal components:
\begin{equation}
T_{\mu\nu} = \frac{1}{32\pi} \begin{pmatrix}
-\dfrac{\omega^2}{r^2} & 0 & 0 & 0 \\
0 & -\dfrac{\omega^2}{r^2} & 0 & 0 \\
0 & 0 & 3\omega^2(1 + \omega^2) & \dfrac{\omega(3\omega^2 + 2)}{r} \\
0 & 0 & \dfrac{\omega(3\omega^2 + 2)}{r} & \dfrac{3\omega^2}{r^2}
\end{pmatrix}.
\end{equation}

The Weak Energy Condition (WEC) requires $T_{\mu\nu}\,U^\mu\,U^\nu \geq 0$ all time-like vectors $U^\mu$. However, this condition is violated near the axis, where the energy density becomes negative:
\begin{equation}
\rho = T_{tt} = -\frac{\omega^2}{32\pi r^2} < 0.
\end{equation}

The principal pressures are given by:
\begin{align}
P_r &= T_{rr} = -\frac{\omega^2}{32\pi r^2}, \\
P_\phi &= T_{\phi\phi} = \frac{3\omega^2(1+\omega^2)}{32\pi}, \\
P_z &= T_{zz} = \frac{3\omega^2}{32\pi r^2}.
\end{align}

The stress tensor is anisotropic, since $P_\phi \neq P_z$ and includes an off-diagonal component:
\begin{equation}
T_{\phi z} = \frac{\omega(3\omega^2 + 2)}{32\pi r},
\end{equation}
which represents a momentum flux between the angular and axial directions, induced by the helical structure of the spacetime.

In the following sections of this paper, we will investigate the effects of this geometry and energy content on geodesic motion and wave optic modes.

\section{Geodesic Dynamics and Helical Trajectories}
\label{sec:geodesics}

In GR, the motion of a free test particle is described by the
geodesic equation
\begin{equation}
\frac{d^{2}x^{\mu}}{d\tau^{2}}
   +\Gamma^{\mu}_{\ \alpha\beta}\,
    \frac{dx^{\alpha}}{d\tau}\,
    \frac{dx^{\beta}}{d\tau}=0,
\label{eq:geodesic-general}
\end{equation}
where $\tau$ is an affine parameter (proper time for time-like curves) and
$\Gamma^{\mu}_{\ \alpha\beta}$ are the Christoffel symbols constructed from
the metric~\eqref{metric}.  Restricting attention to the spatial sector
$\{r,\phi ,z\}$ of the helically twisted geometry and inserting the
non-vanishing symbols, Eq.~\eqref{eq:geodesic-general} splits into the following three second-order equations:
\begin{align}
\ddot r &=
      r\,(1+\omega^{2})\,\dot{\phi}^{2}
    + 2\omega\,\dot{\phi}\,\dot z
    + \omega^{2}\,\dot z^{2},
\label{eq:geod-r}\\[4pt]
\ddot{\phi} &=
      -\frac{2}{r}\Bigl[
         \bigl(1+\tfrac{\omega^{2}}{2}\bigr)\dot r\,\dot\phi
         +\tfrac{\omega}{2}\,\dot r\,\dot z\Bigr],
\label{eq:geod-phi}\\[4pt]
\ddot z &=
      -\omega\,(1+\omega^{2})\,\dot r\,\dot\phi
      -\frac{\omega^{2}}{r}\,\dot r\,\dot z .
\label{eq:geod-z}
\end{align}

Because the metric coefficients do not depend on the coordinates
$t$, $\phi$ or $z$, the vectors
$\partial_{t}$, $\partial_{\phi}$ and $\partial_{z}$ are Killing vectors.
Projecting the four-velocity on each Killing vector produces three integrals
of motion that we denote, respectively, by the conserved energy $\mathrm{E}$, the
azimuthal angular momentum $\mathrm{L}$ and the axial momentum $p_{z}$:
\begin{align}
\mathrm{E} &\equiv -p_{t}= -g_{tt}\,\dot t = \dot t,
\label{eq:E-const}\\
\mathrm{L} &\equiv p_{\phi}
      = g_{\phi\phi}\,\dot\phi + g_{\phi z}\,\dot z,
\label{eq:L-const}\\
p_{z} &= g_{zz}\,\dot z + g_{z\phi}\,\dot\phi.
\label{eq:pz-const}
\end{align}
Exactly as will be seen later in the Lagrangian and Hamiltonian formalisms,
Eqs.~\eqref{eq:E-const}–\eqref{eq:pz-const} supply the first integrals that
render the system effectively one-dimensional.

Contracting the four-velocity with the metric yields the normalization
condition
$g_{\mu\nu}\dot x^{\mu}\dot x^{\nu}=-\epsilon$,
with $\epsilon=1$ for time-like particles and $\epsilon=0$ for null rays.
Eliminating $\dot t$, $\dot\phi$ and $\dot z$ by means of
Eqs.~\eqref{eq:E-const}–\eqref{eq:pz-const} one obtains a single
first-order equation for $r(\tau)$,
\begin{equation}
\dot r^{\,2}+V_{\!\mathrm{eff}}(r)=E^{2},
\label{eq:radial-first-integral}
\end{equation}
where the effective potential
\begin{equation}
V_{\!\mathrm{eff}}(r)=
   -\epsilon
   +\frac{\mathrm{L}^{2}}{r^{2}}
   +(1+\omega^{2})\,p_{z}^{2}
   -\frac{2\,\omega\,\mathrm{L}\,p_{z}}{r}
\label{eq:Veff}
\end{equation}
encodes the entire influence of the helical torsion on the radial motion.  Circular orbits satisfy
$dV_{\!\mathrm{eff}}/dr=0$, and their stability is fixed by the sign of
$d^{2}V_{\!\mathrm{eff}}/dr^{2}$; both features will be analysed in
Sec.~\ref{sec:circular-orbits}.

\begin{figure}[ht!]
\centering
\includegraphics[width=0.47\textwidth]{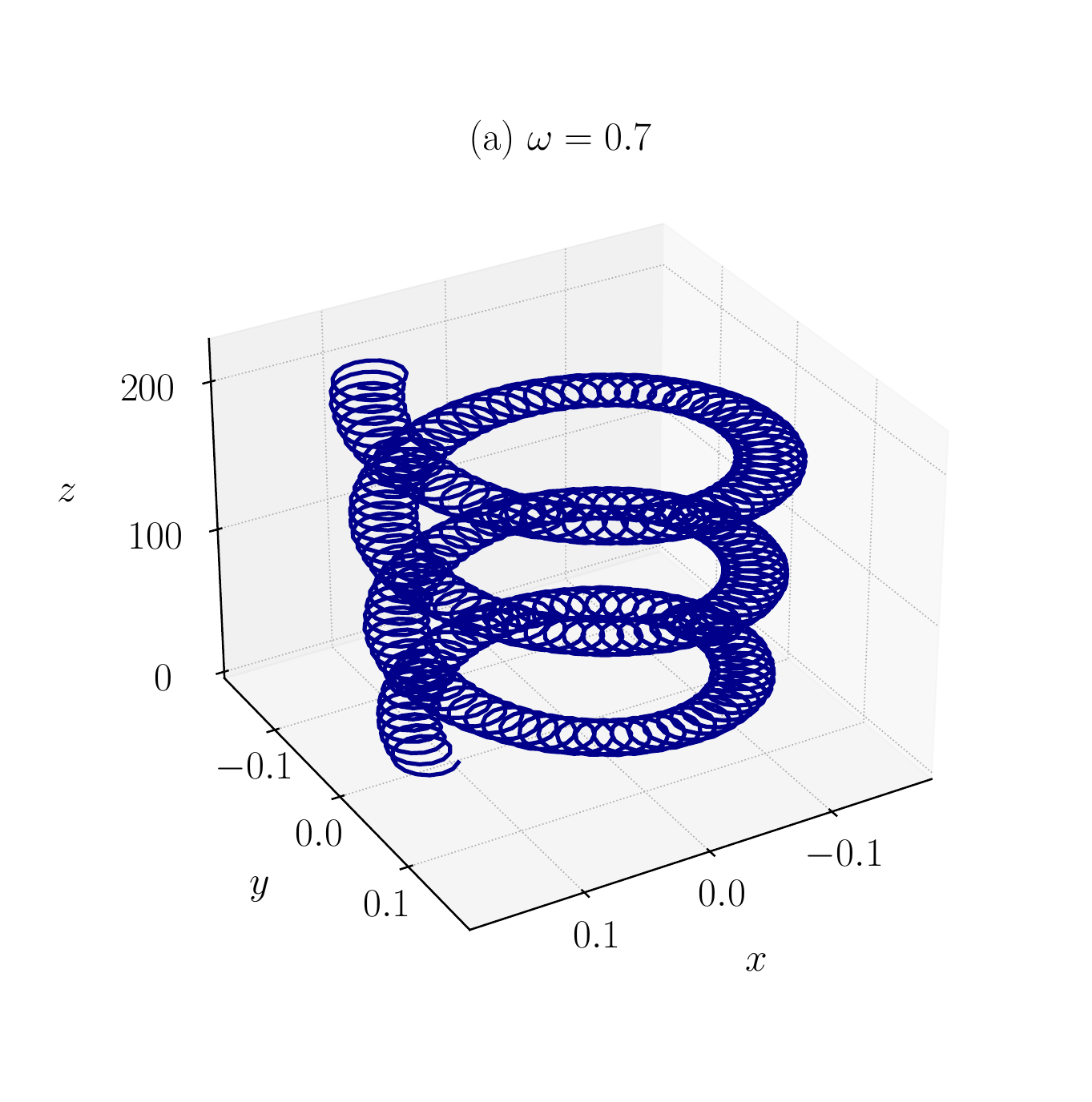}
\includegraphics[width=0.47\textwidth]{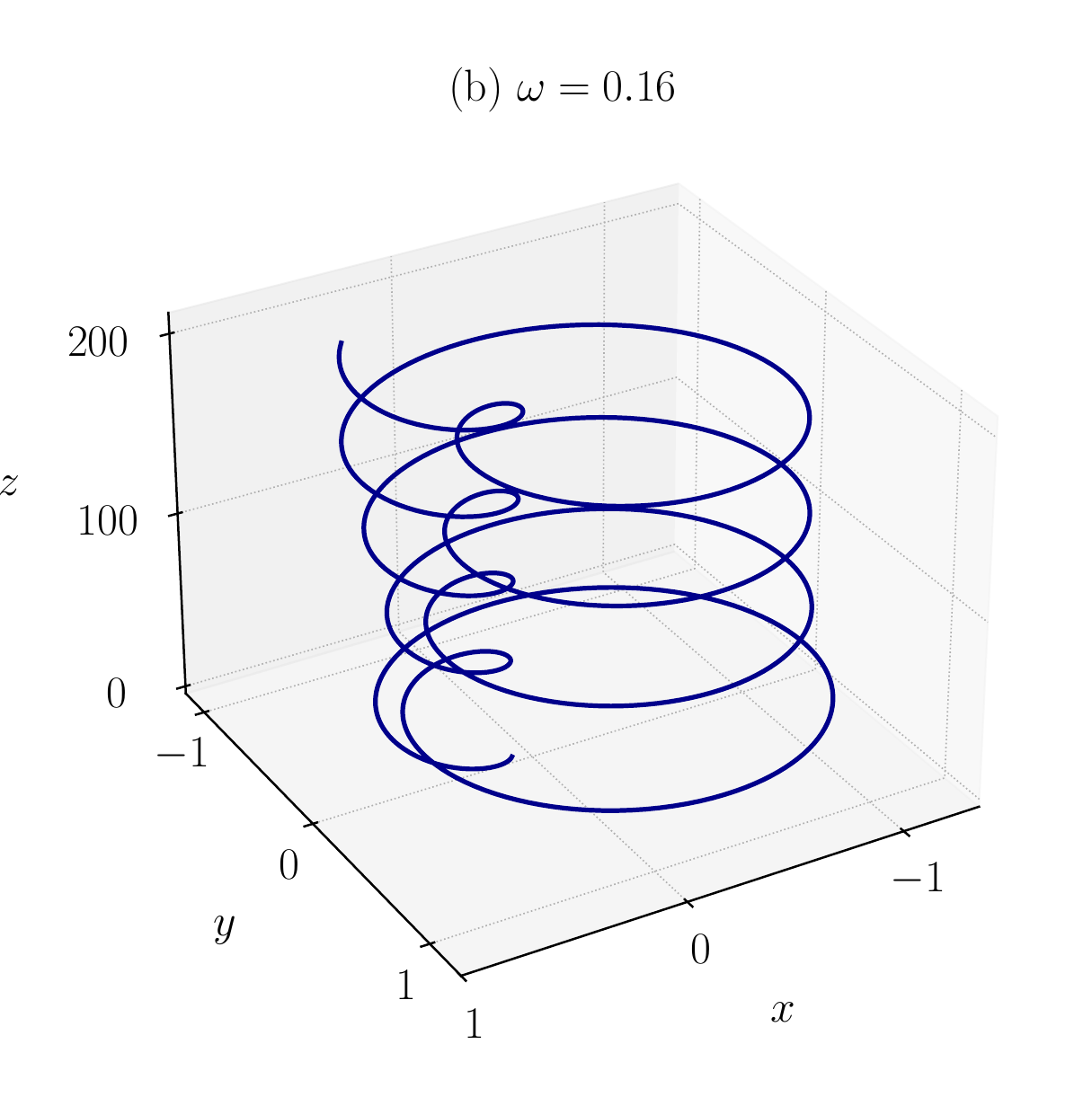}
\caption{\footnotesize Geodesic trajectories in a helically twisted spacetime with torsion. (a) For $\omega = 0.7$, the evolution parameter spans $\tau \in [0, 200]$, and the initial conditions are $ r(0) = 0.1 $, $ \phi(0) = 0.1 $, $z(0) = 0$, $\dot{r}(0) = 0.1$, $\dot{\phi}(0) = 1.0$, and $\dot{z}(0) = 1.0$. (b) For $\omega = 0.16$, with the same evolution interval $ \tau \in [0, 200] $ and the same initial configuration.}
\label{fig:helical_geodesics_panels}
\end{figure}

The same dynamical content can be reformulated starting from the
relativistic Lagrangian
$\mathcal L=\tfrac12\,g_{\mu\nu}\dot x^{\mu}\dot x^{\nu}$.
Because $\mathcal L$ is independent of $t$, $\phi$ and $z$, the associated
Euler–Lagrange equations reproduce the constants of motion
\eqref{eq:E-const}–\eqref{eq:pz-const} through the standard Noether theorem.
A Legendre transform then yields the canonical Hamiltonian
$\mathcal H=\tfrac12\,g^{\mu\nu}p_{\mu}p_{\nu}$,
whose Hamilton equations are equivalent to
Eqs.~\eqref{eq:geod-r}–\eqref{eq:geod-z} and provide the natural setting for the phase-space analysis and the Poincaré maps presented later in
Sec.~\ref{sec:hamiltonian}.  Consequently, the results derived in the Lagrangian and Hamiltonian approaches constitute alternative, but fully
equivalent, descriptions of the trajectories already contained in the basic geodesic equation~\eqref{eq:geodesic-general}.

Figures~\ref{fig:helical_geodesics_panels} and~\ref{fig:geodesics_comparison} illustrate this behavior. They show the particle's spatial evolution for various torsion strengths and initial configurations, demonstrating how geometric torsion tightens the helical structure and reduces the axial pitch as $ \omega $ increases.

Figure~\ref{fig:helical_geodesics_panels} presents two examples of geodesic motion for the same initial configuration but different values of $ \omega $. In panel~(a), for $\omega = 0.7$, the trajectory completes approximately four full rotations along the $z$-axis, with a mean axial step per turn of $\Delta z \approx -53.11,\; \Delta z/\Delta\phi \approx -8.45$. In panel~(b), the curve undergoes five full revolutions using $\omega = 0.16$, with a reduced pitch of $\Delta z \approx -39.10, \; \Delta z/\Delta\phi \approx -6.22$. These results demonstrate how increasing torsion tightens the helix and decreases the vertical propagation, emphasizing the geometric control over particle motion.

Figure~\ref{fig:geodesics_comparison} shows a second set of geodesic curves, where the initial conditions and integration interval vary across panels. In panel~(a), the resulting path forms a tight helix with several revolutions but reduced vertical progression.
In panel~(b), the geodesic forms a more open spiral with much greater axial displacement. These contrasting behaviors further illustrate the impact of the torsion parameter on the structure and inclination of geodesic trajectories.

\begin{figure}[ht!]
\centering
\includegraphics[width=0.47\textwidth]{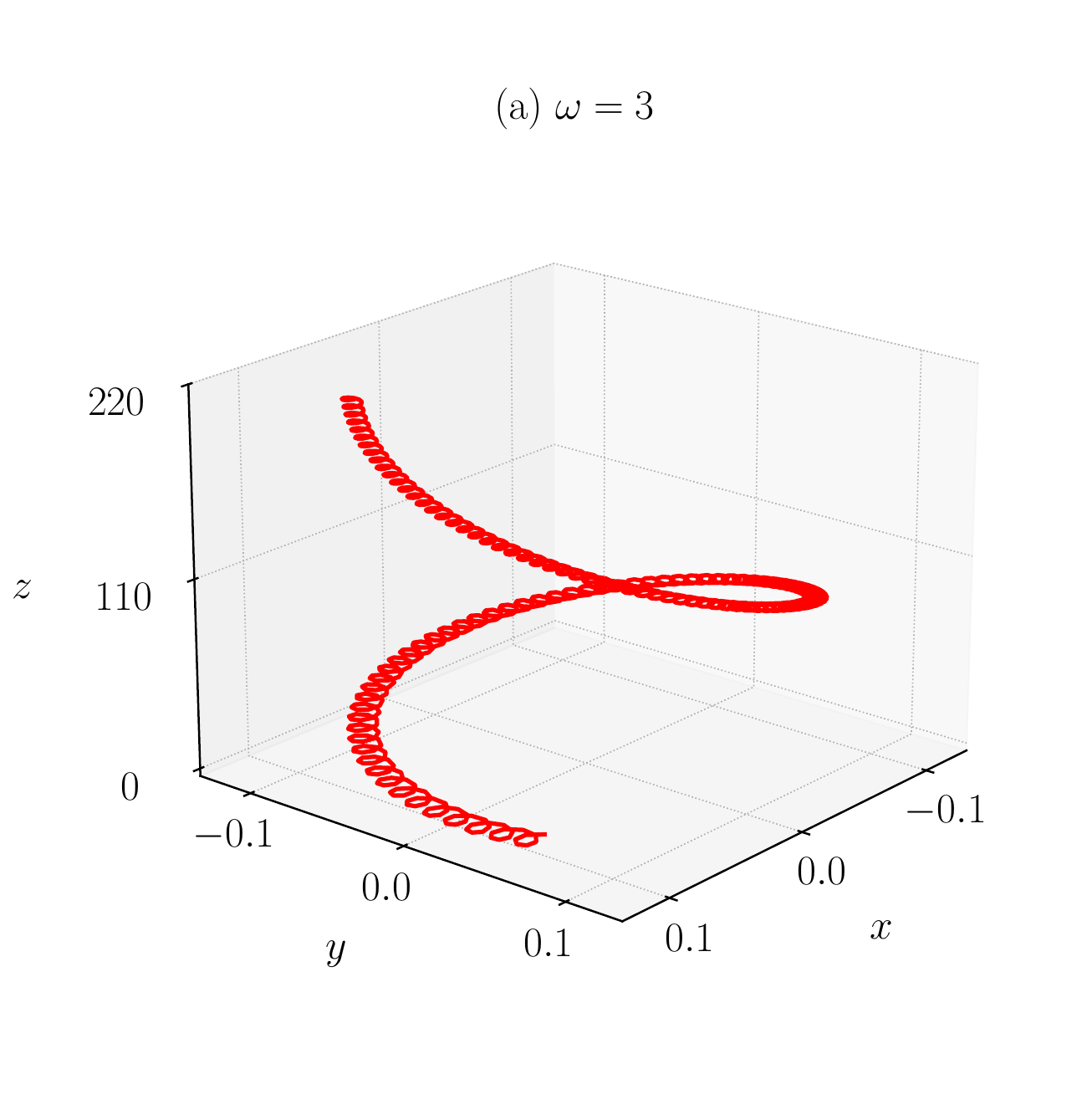}
\includegraphics[width=0.47\textwidth]{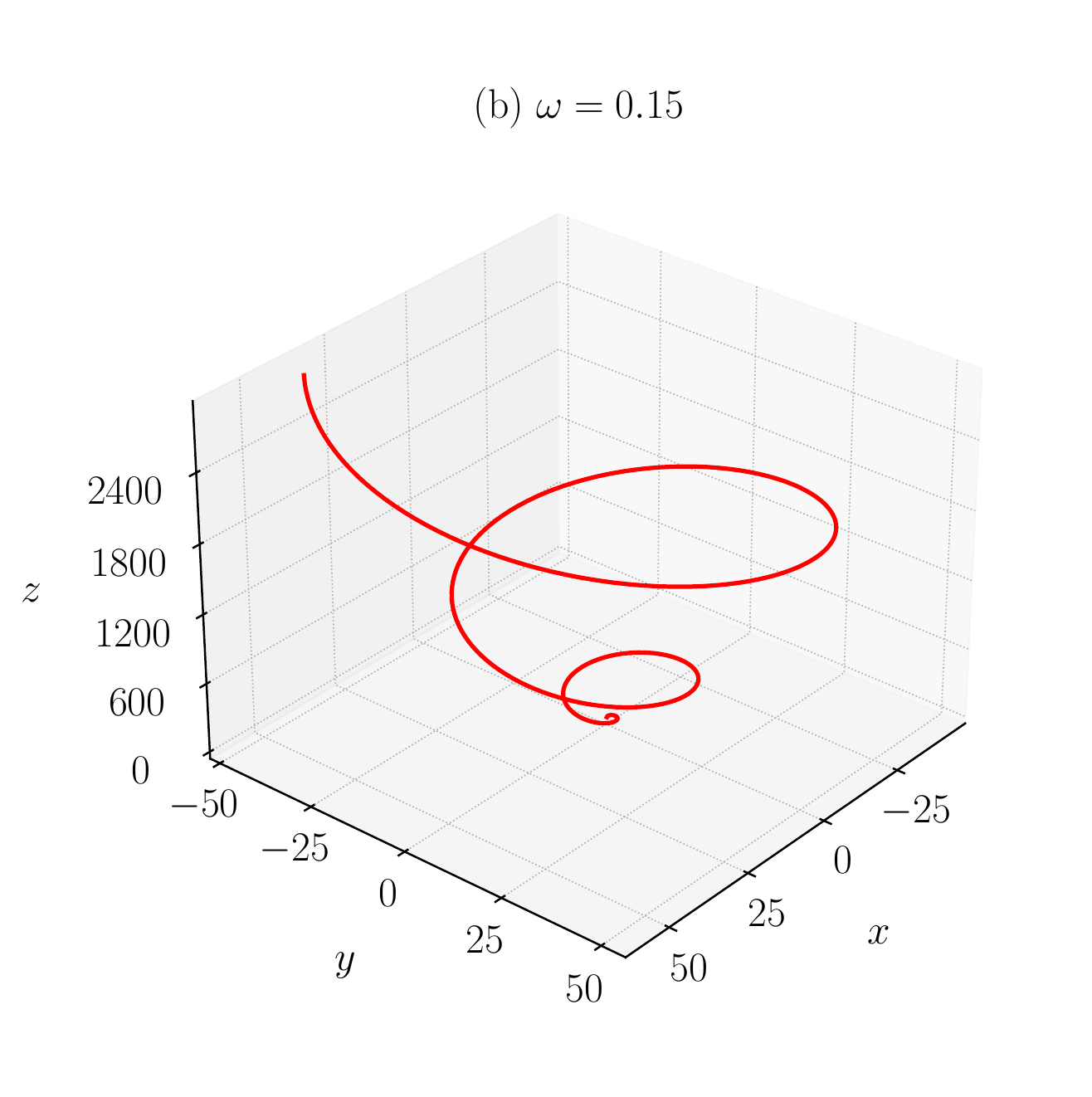}
\caption{\footnotesize Geodesic motion in helicoidal spacetime with torsion for two values of the angular parameter $ \omega $. (a) For $ \omega = 3 $, the evolution parameter spans the interval $ \tau \in [0, 300] $, and the initial conditions are
$ r(0) = 0.1 $, $ \phi(0) = 0.5 $, $ z(0) = 0 $, $ \dot{r}(0) = 0.01 $, $ \dot{\phi}(0) = 1.5 $, and $ \dot{z}(0) = 0.2 $. 
(b) For $ \omega = 0.15 $, with evolution parameter in $ \tau \in [0, 800] $, and initial conditions
$ r(0) = 0.2 $, $ \phi(0) = 2.7 $, $ z(0) = 0 $, $ \dot{r}(0) = 0.01 $, $ \dot{\phi}(0) = 2.7 $, and $ \dot{z}(0) = 3.5 $.}
\label{fig:geodesics_comparison}
\end{figure}

\begin{figure*}[ht!]
\centering
\includegraphics[width=0.47\textwidth]{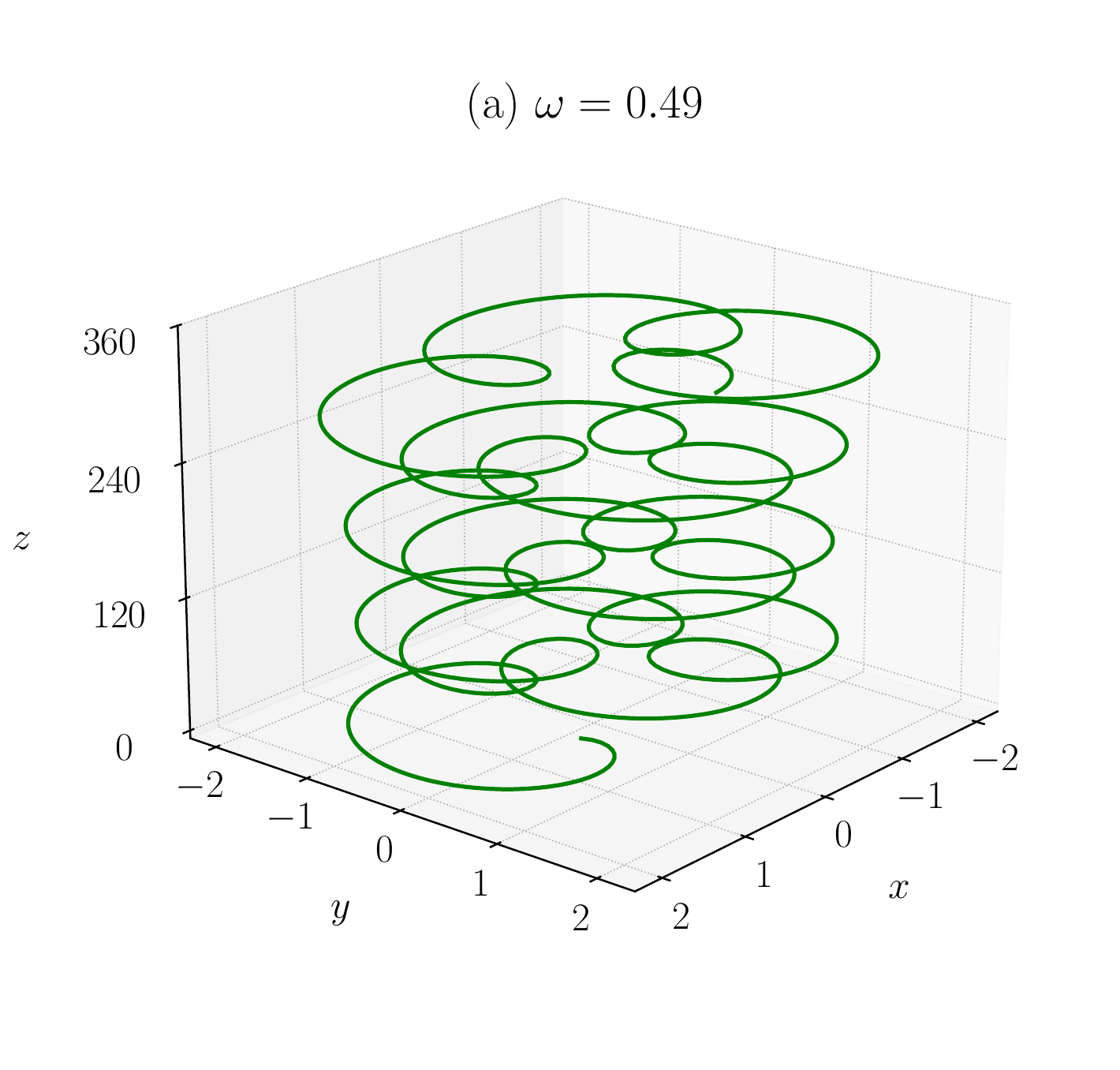}\qquad    \includegraphics[width=0.47\textwidth]{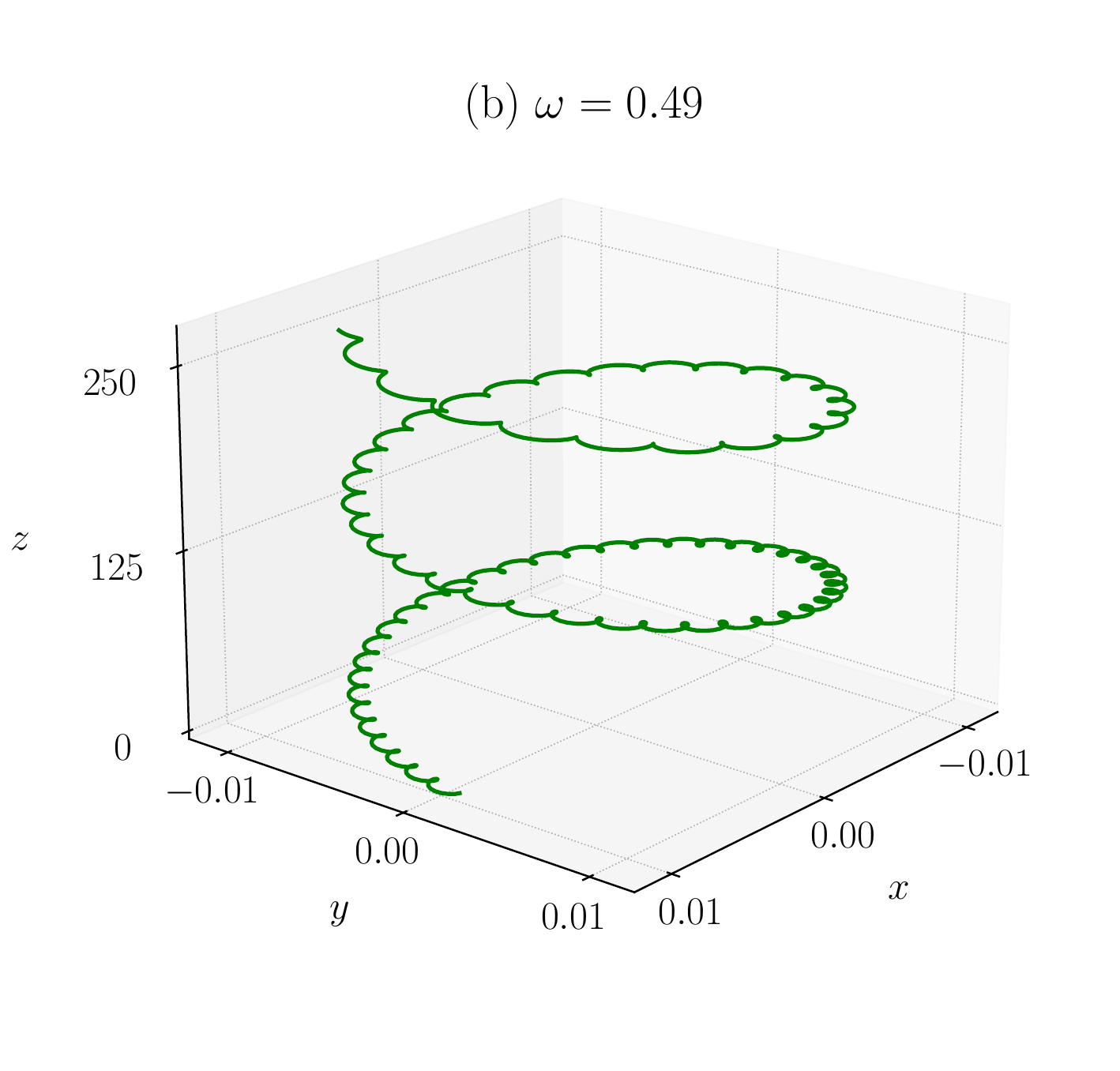}
\includegraphics[width=0.47\textwidth]{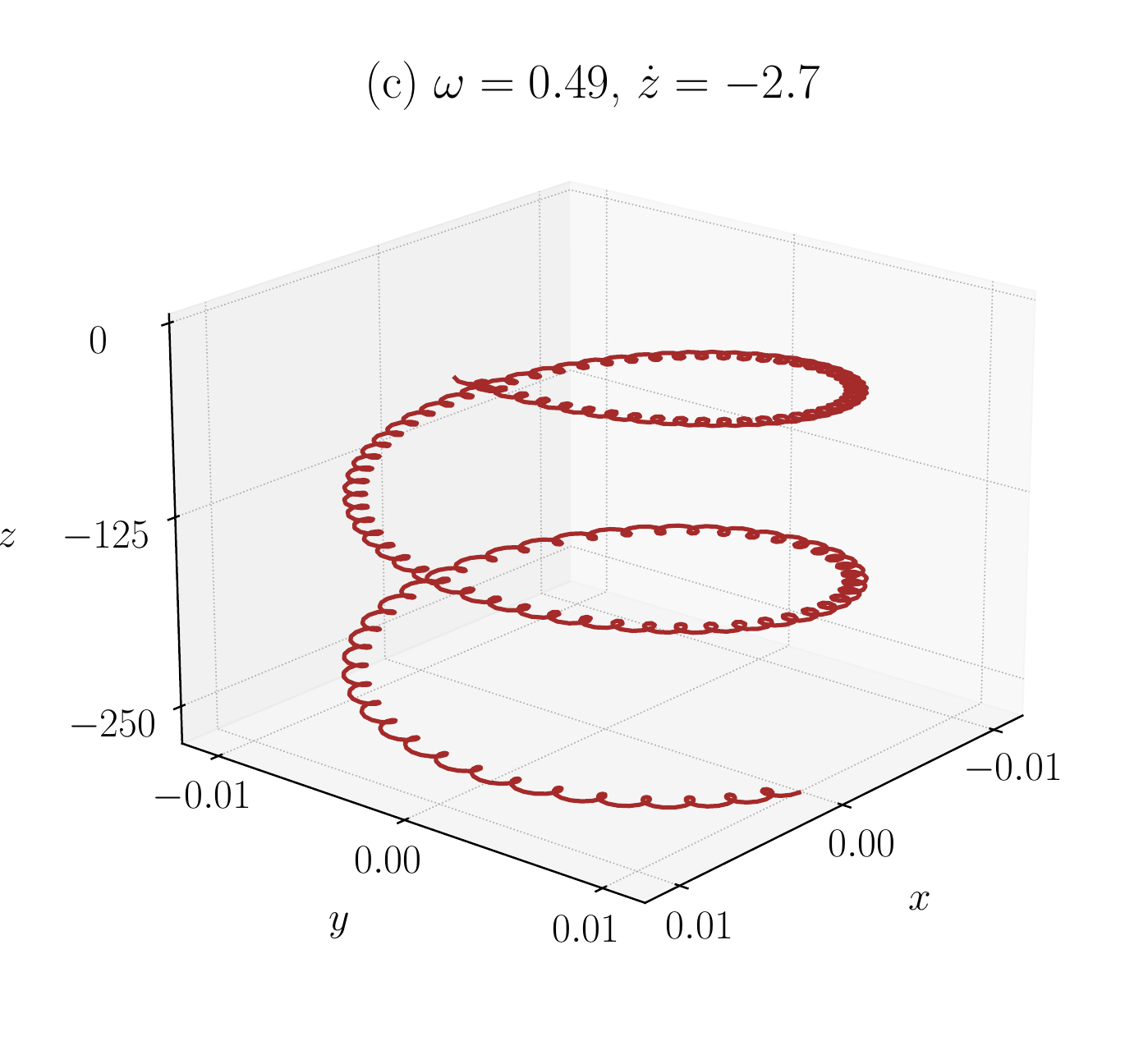} 
\caption{\footnotesize Geodesic trajectories in the helicoidal spacetime for fixed angular parameter $\omega = 0.49$. In panel (a), the initial conditions are $r(0) = 0.5$, $\phi(0) = 0.5$, $z(0) = 0$, $\dot{r}(0) = 0.05$, $\dot{\phi}(0) = 1.9$, and $\dot{z}(0) = 2.7$. In panel (b), the trajectory evolves from a region much closer to the axis, with $r(0) = 0.01$, and other initial conditions as $\phi(0) = 0.1$, $z(0) = 0$, $\dot{r}(0) = 0.05$, $\dot{\phi}(0) = 1.9$, and $\dot{z}(0) = 2.7$. In both cases, the evolution parameter runs in the interval $\tau \in [0, 100]$. In panel (c), the same initial radial and angular values of panel (b) are considered, but with a negative vertical velocity, $\dot{z}(0) = -2.7$ (i.e., $p_z<0$).} 
\label{fig:geodesics_w049}
\end{figure*}

Figure~\ref{fig:geodesics_w049} displays three representative geodesic trajectories in the helicoidal spacetime for the fixed value $\omega = 0.49$. In panel (a), the particle starts at a moderate radial distance and follows a helical path with significant winding and vertical displacement. In contrast, panel (b) shows the behavior of a particle initialized very close to the axis of symmetry. Due to the strong influence of the torsional geometry near the origin, the trajectory exhibits a pronounced spiral structure, with tighter coils and stronger coupling between the azimuthal and axial motion. Panel (c) explores the effect of reversing the vertical velocity component while keeping the other initial conditions of panel (b) unchanged. This is explained by the reversal of the coupling term $\omega\mathrm{L}p_{z}$ in Eq.~\eqref{eq:Veff}, which occurs due to the condition $\dot{z} < 0$ (see Eq.~\eqref{eq:pz-const}).
As a result, the trajectory develops a downward helical motion, still confined near the axis initially, but now propagating in the negative $z$-direction. Interestingly, as the particle descends, the radius of the helix gradually increases, reflecting a transfer of axial motion into radial displacement induced by the torsion term. This illustrates how the axial velocity sign modifies the geodesic's global orientation and shape while preserving the torsion-induced spiral structure. These results highlight both the sensitivity of the geodesic structure to the initial radial position and the influence of the direction and coupling of motion in the axial coordinate in the presence of torsion. A similar effect of torsion (dislocation) on geodesic dynamics was observed in Ref.~\cite{dos2024dynamics}.

\section{Special Orbits and Stability} \label{sec:hamiltonian}

We also adopt the Hamiltonian formalism to investigate the classical motion of test particles in the helically twisted spacetime with torsion. This approach allows us to derive the full set of dynamical equations and facilitates a systematic study of stability through linear analysis and phase space behavior.

We start with the classical Lagrangian derived from the metric (\ref{metric}) given by
\begin{equation}
\mathcal{L} = \frac{1}{2} \left[ -\dot{t}^2 + \dot{r}^2 + r^2 \dot{\phi}^2 + (\omega\, r\, \dot{\phi} + \dot{z})^2 \right],\label{lag}
\end{equation}
where the dot denotes a derivative with respect to the affine parameter $\tau$.

The Legendre transformation gives the Hamiltonian
\begin{equation}
\mathcal{H} = p_t\, \dot{t} + p_r\, \dot{r} + p_\phi\, \dot{\phi} + p_z\, \dot{z} - \mathcal{L}.
\end{equation}

Expressed in terms of the canonical momenta using Eqs. (\ref{eq:E-const})-(\ref{eq:pz-const}), the Hamiltonian becomes
\begin{align}
\mathcal{H} &= \frac{1}{2} \left[ p_r^2 - p_t^2 + \frac{(\omega^2 + 1)^2}{r^2} (\omega p_z r - p_\phi)^2 \right]\notag \\& + \frac{1}{2}\left(\frac{p_z - \omega p_\phi/r}{\omega^2 + 1} \right)^2.
\end{align}

The equations of motion follow from Hamilton's canonical equations:
\begin{equation}
\dot{x}^\mu = \frac{\partial \mathcal{H}}{\partial p_\mu}, \quad \dot{p}^\mu = -\frac{\partial \mathcal{H}}{\partial x^\mu}.
\end{equation}
Explicitly, $\dot{x}^\mu$ and $\dot{p}^\mu$ are expressed as
\begin{align}
\dot{t} &= -p_t, \\
\dot{r} &= p_r, \\
\dot{\phi} &= \frac{-2\omega^3 p_z r + 2\omega^2 p_\phi - \omega p_z r + p_\phi}{r^2(\omega^4 + 2\omega^2 + 1)}, \\
\dot{z} &= \frac{3\omega^4 p_z r - 2\omega^3 p_\phi + 3\omega^2 p_z r - \omega p_\phi + p_z r}{r(\omega^4 + 2\omega^2 + 1)}.
\end{align}

The conjugate momenta evolve as
\begin{align}
\dot{p}^t &= 0, \\
\dot{p}^r &= \frac{p_\phi (-2\omega^3 p_z r + 2\omega^2 p_\phi - \omega p_z r + p_\phi)}{r^3(\omega^4 + 2\omega^2 + 1)}, \\
\dot{p}_\phi &= 0, \\
\dot{p}_z &= 0.
\end{align}

In the following, we will use this set of equations to study orbital stability and circular motion conditions.

\subsection{Poincaré Maps and Stability}

Now we examine the stability of the orbits.  A linear perturbation analysis is carried out by expanding the equations of motion around a reference solution $X(\tau)$ \cite{Poincare1892,LichtenbergLieberman1992,Wiggins2003, Ott2002}.  To assess the stability of a given orbit, we linearise the Hamiltonian equations around a reference solution $X(\tau)$. We collect the canonical variables in the phase-space vector
\begin{equation}
X=(r,\;\phi,\;z,\;p_r,\;p_\phi,\;p_z)^{\mathsf T},
\label{eq:vector-X}
\end{equation}
where the overdot denotes differentiation with respect to the affine parameter $\tau$.  In the helicoidal metric, the full set of equations of motion derived from the Hamiltonian approach described above reads
\begin{align}
\dot r &= p_r, \label{eq:ham-full-a}\\
\dot\phi &=
 \frac{p_\phi-\omega r p_z}{D\,r^{2}},  \label{eq:ham-full-b}\\
\dot z &=
 \frac{(3\omega^{2}+1)p_z}{D}
 -\frac{\omega(2\omega^{2}+1)p_\phi}{D^{2}\,r}, \label{eq:ham-full-c}\\
\dot p_r &=
 \frac{(2\omega^{2}+1)\bigl(\omega r p_z-p_\phi\bigr)p_\phi}{D^{2}\,r^{3}},
 \label{eq:ham-full-d}\\
\dot p_\phi &= 0,\label{eq:ham-full-e}\\
\dot p_z &= 0,\label{eq:ham-full-f}
\end{align}
with the shorthand $D\equiv1+\omega^{2}$. The last two equalities reproduce the integrals of motion $L=p_\phi$ and $p_z$.

Let $X(\tau)$ solve Eqs. (\ref{eq:ham-full-a})-(\ref{eq:ham-full-f}) exactly and introduce a small deviation $\delta X(\tau)$.  Expanding to first order gives
\begin{equation}
\delta\dot X = J(\tau)\,\delta X,
\qquad
J(\tau)\equiv\qty(\pdv{\dot X}{X})_{X(\tau)},
\label{eq:jacobian-def}
\end{equation}
where the $6\times6$ matrix $J$ is the Jacobian of the full evaluated
along the trajectory \cite{GuckenheimerHolmes1983,Wiggins2003}.  Because $\dot p_\phi=\dot p_z=0$,
$J$ acquires a block form
$
J=\begin{psmallmatrix}
\mathbf 0 & \mathbf 0\\[2pt]
\mathbf A & \mathbf B
\end{psmallmatrix},
$
with the non-vanishing entries
\begin{align}
\mathbf A &=
\begin{pmatrix}
0 & 0 & 0\\[4pt]
\displaystyle
\frac{2(p_\phi-\omega r p_z)}{D\,r^{3}} & 0 & 0\\[10pt]
\displaystyle
-\frac{(3\omega^{2}+1)p_z}{D\,r^{2}} & 0 & 0
\end{pmatrix}, 
\mathbf B =
\begin{pmatrix}
1 &
 0 &
 0\\[6pt]
0 &
 \dfrac{1}{D\,r^{2}} &
 -\dfrac{\omega}{D\,r}\\[14pt]
0 &
 -\dfrac{\omega(2\omega^{2}+1)}{D^{2}\,r} &
 \dfrac{3\omega^{2}+1}{D}
\end{pmatrix}, \nonumber\\[2pt]
J_{55}&=
\dfrac{(2\omega^{2}+1)(\omega r p_z-2p_\phi)}{D^{2}r^{3}},\qquad
J_{56}= \dfrac{\omega(2\omega^{2}+1)p_\phi}{D^{2}r^{2}},\nonumber\\
J_{51}&=-\dfrac{3(2\omega^{2}+1)(\omega r p_z-p_\phi)p_\phi}{D^{2}r^{4}}.
\label{eq:jacobian-entries}
\end{align}
All remaining elements not shown explicitly are zero.  Therefore, every component of $J$ is an analytic, rational function of $r(\tau)$, the conserved momenta $p_\phi,\,p_z$, and the parameter $\omega$; no numerical approximation is involved in its construction.

Solving the characteristic equation
$\det\!\bigl(J-\lambda\mathbb I\bigr)=0$
provides the six Lyapunov exponents $\lambda_i$ local to the selected point on the orbit \cite{Lyapunov1892,FILIP_2019,Benettin1980,Wolf1985}.  Purely imaginary $\lambda_i$ indicate linear
stability, whereas a real part different from zero signals local instability.  To characterise global behaviour, one follows the evolution of $\delta X$ according to \eqref{eq:jacobian-def} and evaluates
\begin{equation}
\lambda_{\mathrm{max}}
 =\lim_{\tau\to\infty}\frac{1}{\tau}
  \ln\!\frac{\|\delta X(\tau)\|}{\|\delta X(0)\|},
\label{eq:lyapunov}
\end{equation}
which measures the average exponential rate of divergence between neighbouring trajectories.

The phase-space portraits of Fig.~\ref{fig:poincare} are obtained from the numerical integration of the full Hamiltonian $\dot X = F(X;\omega)$ in Eqs.~(\ref{eq:ham-full-a})-(\ref{eq:ham-full-f}). Whenever the trajectory crosses the hypersurface $z=z_0$ with positive vertical velocity ($\dot z>0$), the pair $(r,p_r)$ is recorded; these pairs constitute the Poincaré map. The variational system $\delta\dot X = J(\tau)\,\delta X$ in Eq.~\eqref{eq:jacobian-def} is used solely for local stability diagnostics (eigenvalues and Lyapunov exponents)
and is not involved in generating the map itself.

Here, the orbit $X(\tau)$ is obtained
numerically from (\ref{eq:ham-full-a})-(\ref{eq:ham-full-f});  at each integration step the analytic
expressions \eqref{eq:jacobian-entries} are evaluated to update
$J(\tau)$.  Time-ordered multiplication of the resulting fundamental
matrices yield the finite-time Lyapunov exponent, while Poincaré sections are constructed by recording intersections with a fixed hypersurface ($\dot z>0,\,z=z_{0}$) in the reduced phase space
$(r,p_{r})$ as illustrated in  Fig.~\ref{fig:poincare}.  The combined use of the analytic Jacobian and the numerical trajectory, thus linking the local linear analysis to the global phase-space portraits, provides a consistent picture of how the torsion parameter $\omega$ governs stability and orbital architecture.
\begin{figure}[!t]
\centering
\includegraphics[width=0.48\textwidth]{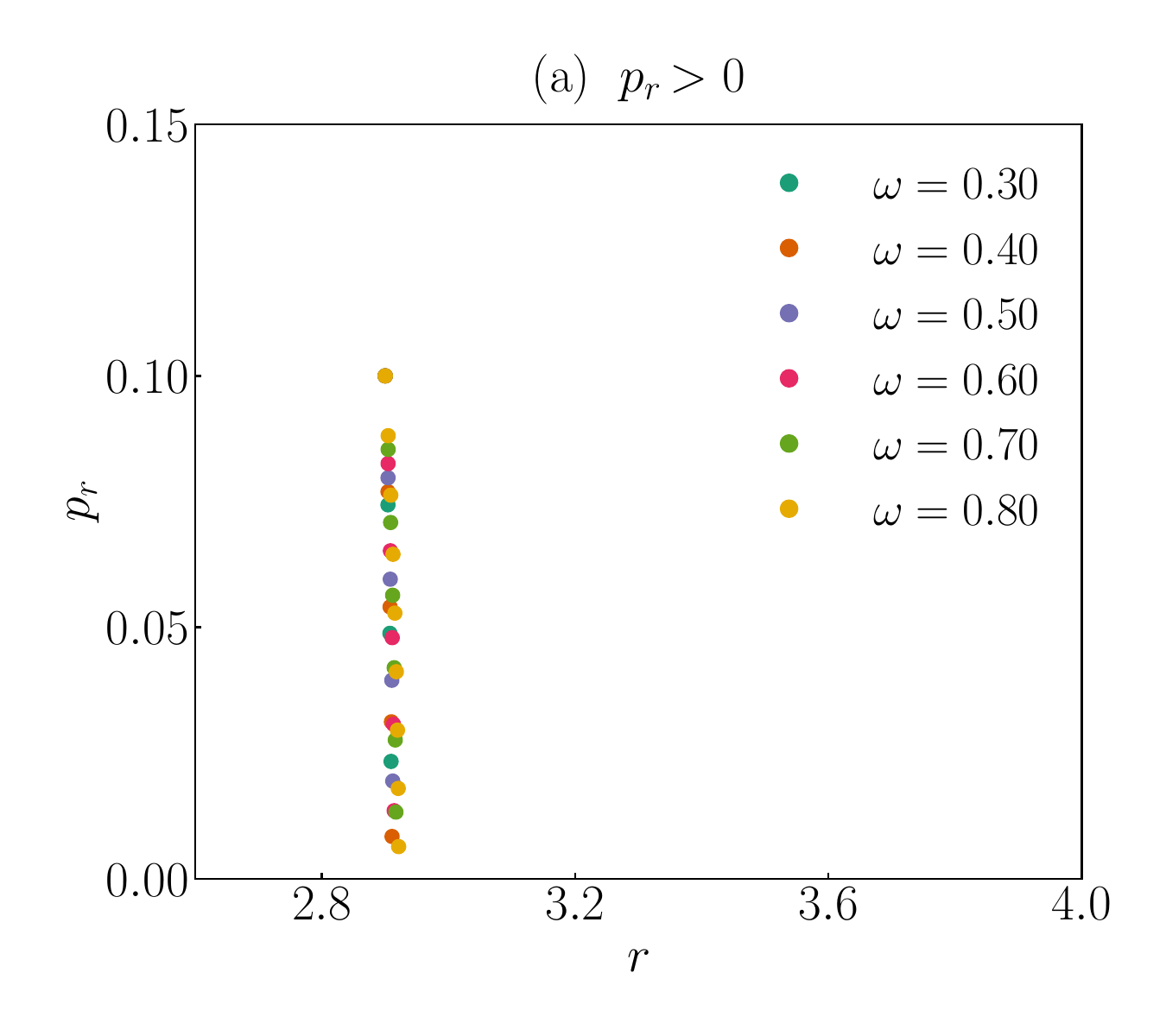}
\hfill
\includegraphics[width=0.48\textwidth]{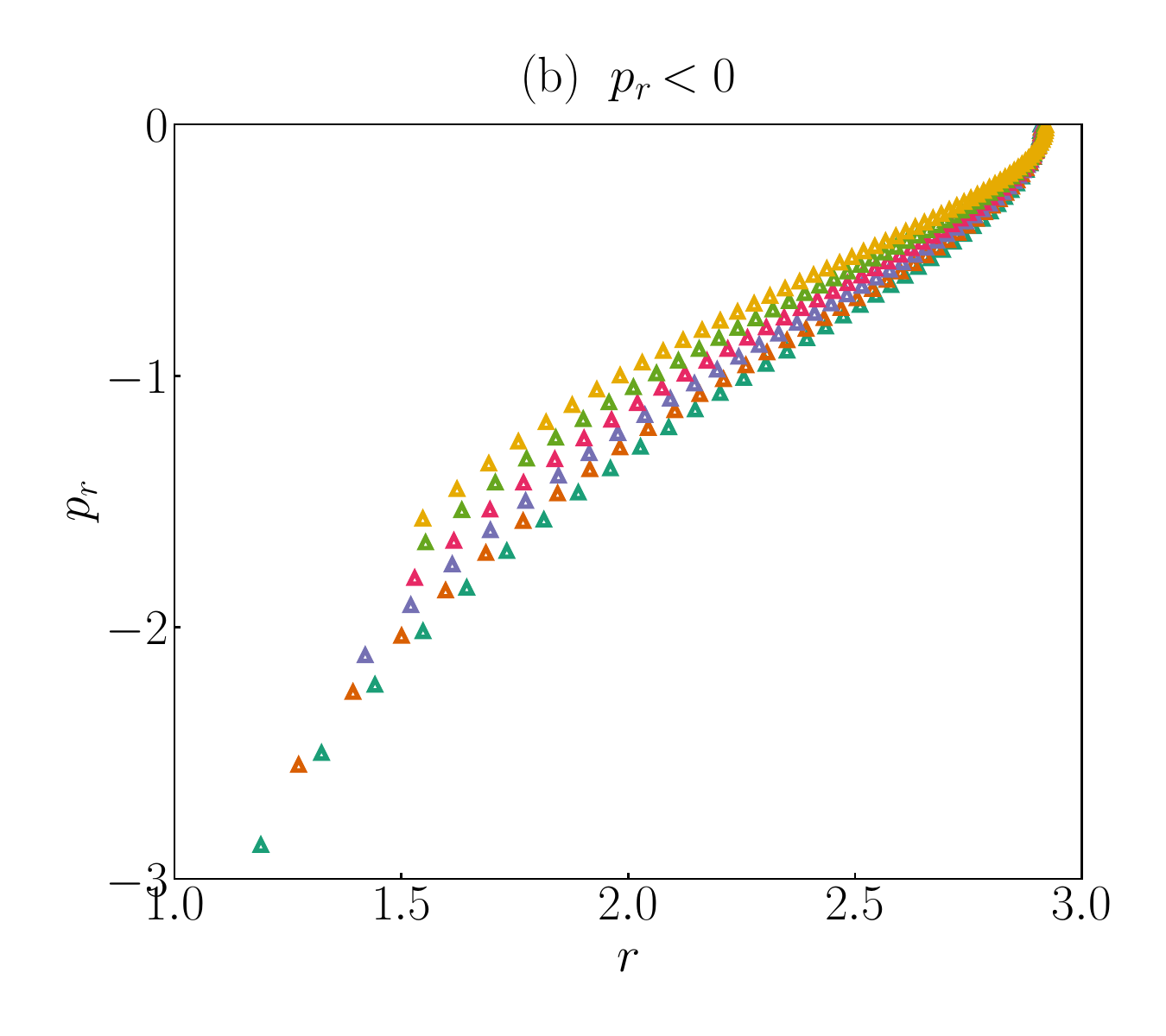}
\caption{\footnotesize Poincaré maps on the plane $z=z_0=2$.
(a)~Outbound crossings ($p_r>0$) collapse into a narrow vertical strip centred at $r\simeq3.0$, indicating that the orbit meets the reference plane near the outer turning point of each revolution. (b)~Inbound crossings ($p_r<0$) populate a much broader band, $1\lesssim r\lesssim3$; the pattern evolves from thin, nearly integrable
curves at small~$\omega$ to wider, quasi-periodic layers as $\omega$
increases.}
\label{fig:poincare}
\end{figure}

\begin{figure}[htbp]
\centering
\includegraphics[width=0.92\linewidth]{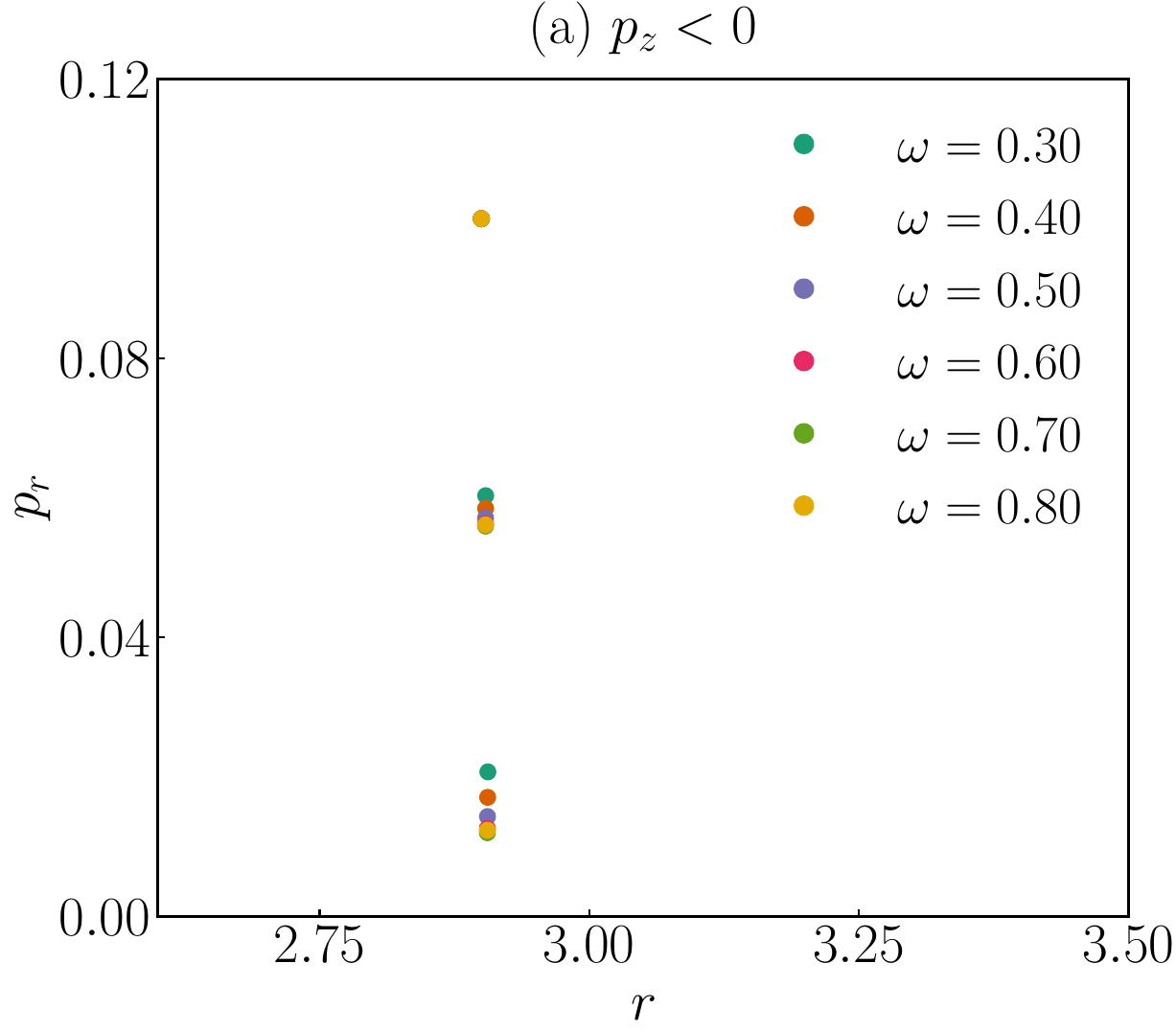}\hfill  \includegraphics[width=0.92\linewidth]{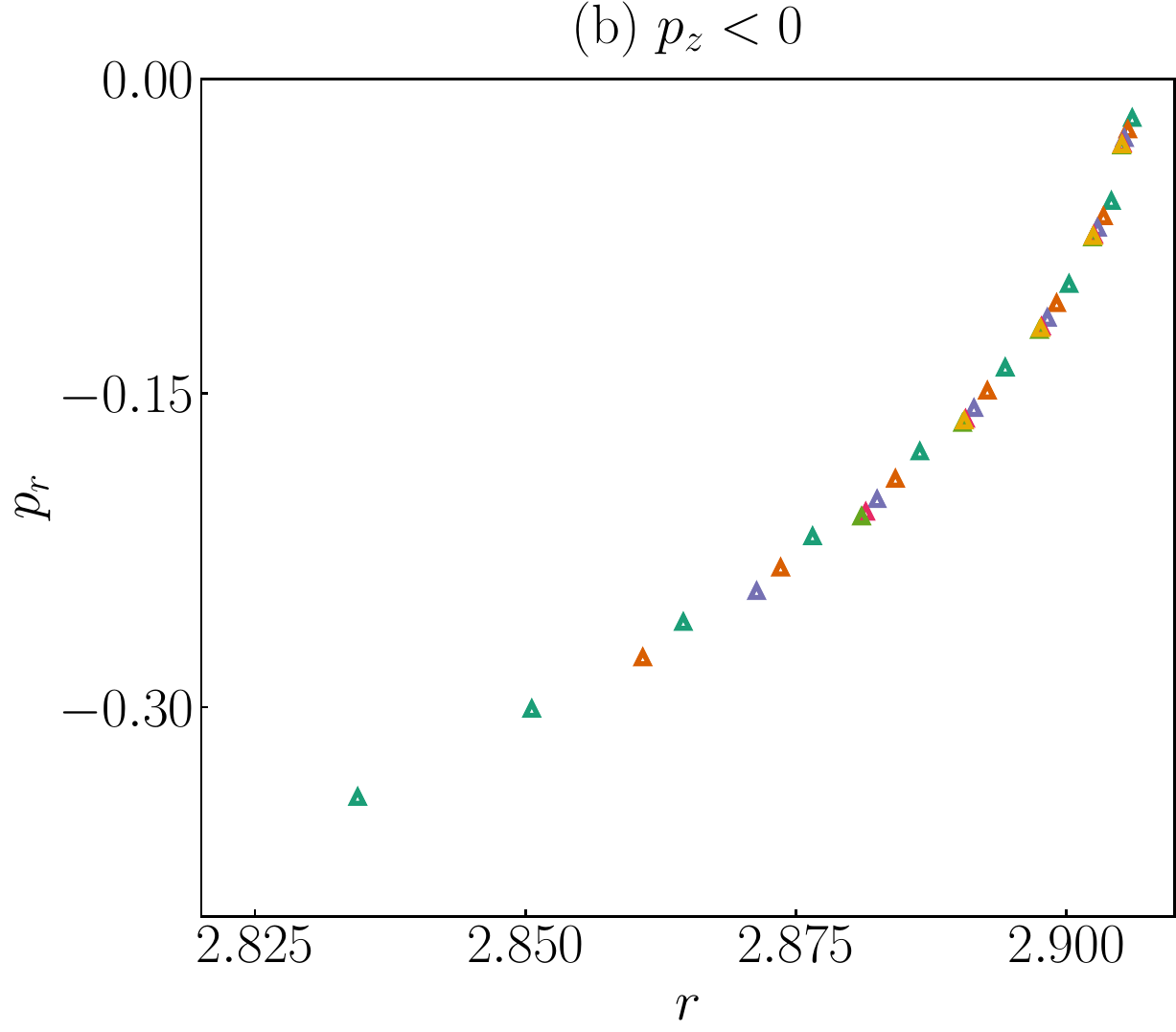}
\caption{\footnotesize
Poincaré sections on the plane $z=z_{0}=2$ for orbits with negative axial momentum $p_{z}<0$.
(a) Outbound crossings ($\dot z<0$, $p_{r}>0$) cluster in a narrow vertical strip around $r\simeq3$, mirroring the behaviour seen for $p_{z}>0$ but shifted in phase because the trajectory meets the reference plane while descending.
(b)~Inbound crossings ($\dot z<0$, $p_{r}<0$) populate a very thin band just below $p_{r}=0$, indicating that the repulsive coupling term $-2\omega L p_{z}/r$ prevents the particle from penetrating deeply into the potential well.
Colours encode the torsion strength $\omega=0.3\text{-}0.8$.
Together with Fig.~\ref{fig:poincare}, these panels show that changing the sign of $p_{z}$ rearranges the phase-space islands without destroying the overall layer structure, confirming that the product $\omega L p_{z}$ is the key control parameter.}
\label{fig:poincare_pzneg}
\end{figure}
\begin{figure}[!t]
\centering
\includegraphics[width=0.9\linewidth]{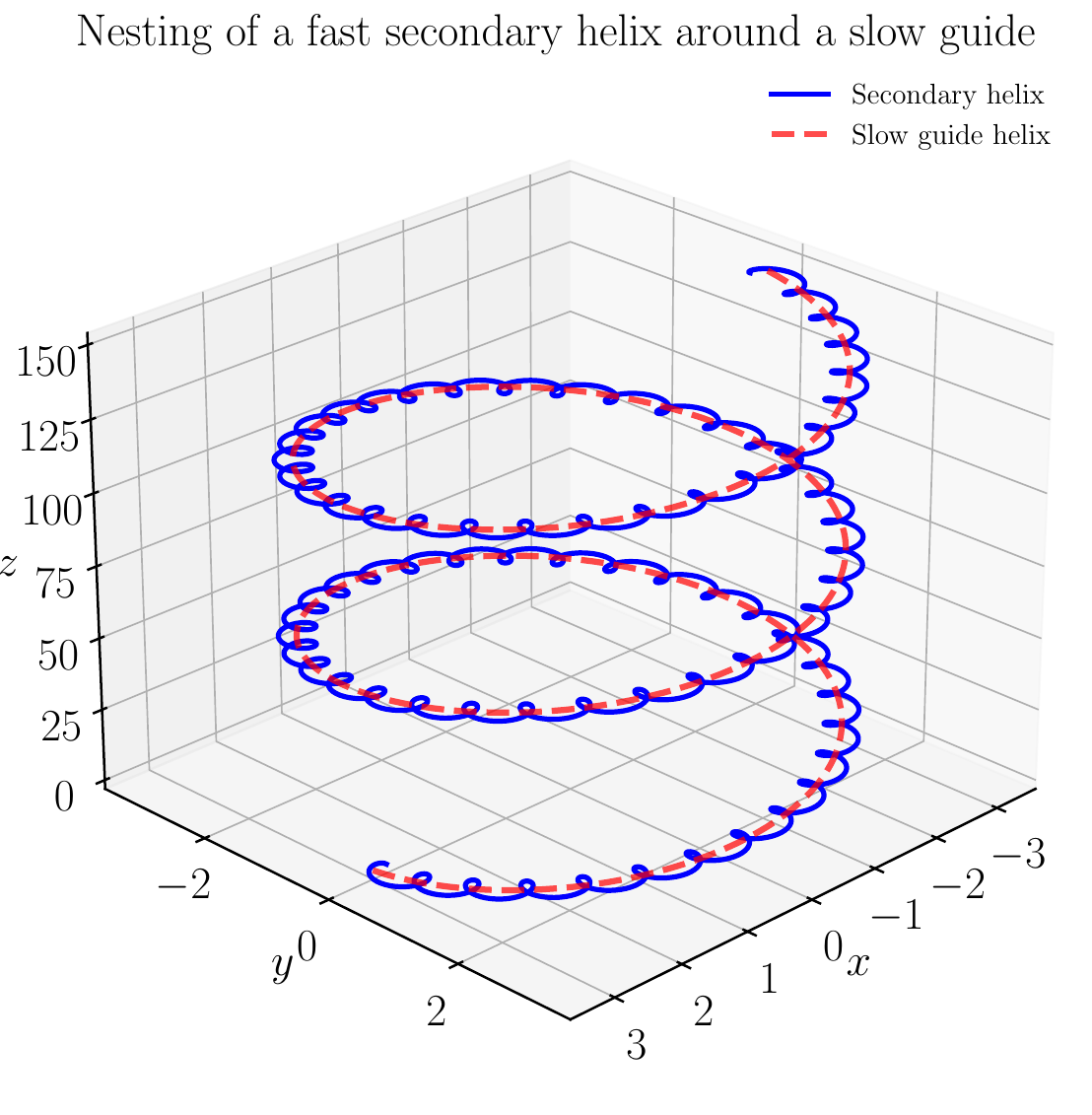}
\caption{\footnotesize
The nested helical orbit is generated directly from the Lagrangian
\eqref{lag}.  The red curve is the slow guide helix $\mathbf G(\tau)=\bigl(r_{0}\cos\Omega_{\!g}\tau, r_{0}\sin\Omega_{\!g}\tau,v_{z}\tau\bigr)$
with $r_{0}=P_{\phi}/(\omega P_{z})$,
$\Omega_{\!g}=\omega^{2}P_{z}/P_{\phi}$ and
$v_{z}=P_{z}/(1+\omega^{2})$.
The blue curve is obtained by adding the circular
offset $A\bigl[\cos(k\tau)\mathbf N+\sin(k\tau)\mathbf B\bigr]$,
where $(\mathbf T,\mathbf N,\mathbf B)$ is the Frenet frame of
$\mathbf G$ \cite{ONeill2006,LandauLifshitz1976,Frankel2012}. Because the perturbation lives entirely in the normal plane, i.e, the distance between the two helices remains constant
secondary radius $A$, realizing an exact ``super-helical’’ motion in
which a fast wave of frequency $k$ rides on the slow drift set by
$\Omega_{\!g}$. We use the parameters $\omega=0.5$, $P_\phi=4$, $P_z=2$,
$A=0.8$, $k=5$.}
\label{fig:nested_superhelix}
\end{figure}

To complement the analytical criteria, the Hamiltonian equations are
integrated numerically, and the resulting motion is visualised through a set
of graphical diagnostics.  First, the complete three-dimensional trajectory is reconstructed in Cartesian coordinates $(x(\tau),y(\tau),z(\tau))$, exposing at a glance the helical character of the path. Two planar projections, $(r,\phi)$ and $(\phi,z)$, are 
extracted to emphasise the radial excursions and the axial advance per revolution.  In parallel, the effective potential $V_{\!\text{eff}}(r)$ from Eq.~\eqref{eq:Veff} is plotted for the same
integrals of motion, allowing a direct comparison between the turning points predicted by the one-dimensional reduction and those realised in the whole orbit. Poincaré sections provide a compact portrait of the phase-space structure. Throughout this work, we adopt the plane $z=z_{0}=2$ and keep only the
upward crossings, $\dot z>0$. For every value of the torsion parameter $\omega$ we integrate the full Hamiltonian (\ref{eq:ham-full-a})-(\ref{eq:ham-full-f}); at each step the crossing criterion is
checked and, whenever it is met, the pair $(r,p_{r})$ is stored. We separate the data according to the sign of the radial momentum to highlight the two radial branches of the motion. Figure~\ref{fig:poincare} collects the resulting maps: The stark contrast between panels (a) and (b) confirms that the helicoidal torsion fixes the phase at which the trajectory re-enters the plane during the outward leg, the inward leg is far more sensitive to $\omega$.
For slight torsion, the points lie on thin invariant curves; increasing $\omega$ broadens those curves into bands, signalling a transition from regular to mixed dynamics. These global features echo the local‐stability diagnostics obtained from the Jacobian analysis, reinforcing the role of $\omega$ as an effective control parameter for the system's orbital architecture.

For completeness, we repeated the construction with the axial momentum reversed. Figure~\ref{fig:poincare_pzneg} displays the resulting sections for $p_{z}<0$, selecting crossings with $\dot z<0$ so that the reference plane is approached from the same physical side. Panel~(a) reveals that the outbound strip persists but is phase-shifted to the descending branch, while panel~(b) shows that the inbound points collapse into a very thin layer close to $p_{r}=0$, a direct consequence of the repulsive sign acquired by the coupling term $-2\omega Lp_{z}/r$. The comparison with Fig.~\ref{fig:poincare} demonstrates that it is the product $\omega Lp_{z}$, rather than the individual signs, that determines the topology of the invariant layers; reversing either $\omega$ or $p_{z}$ relocates the islands without altering their stability hierarchy.

Having established the global picture in phase space, it is natural to ask how sensitive a single timelike orbit is to rapid, small–amplitude
perturbations that respect the conserved quantities
$(\mathrm E,\mathrm L,p_{z})$.
Within the Lagrangian framework, this question can be answered analytically:
starting from the slow helical geodesic of radius
$r_{0}=P_{\phi}/(\omega P_{z})$, we add a transverse oscillation
of fixed amplitude $A$ that rotates $k$ times faster around the local
Frenet-Serret frame \cite{ONeill2006,LandauLifshitz1976,Frankel2012,doCarmo1976,Struik1950}.
The resulting ``helix on a helix’’ is shown in
Fig.~\ref{fig:nested_superhelix} provides a concrete example of
persistent confinement under high–frequency modulation.

The construction confirms, at the level of explicit trajectories, the
Jacobian-based result that torsion does not necessarily disrupt bounded
motion: Even under rapid radial wobbling, the particle never escapes the
effective potential well, its closest approach to the axis being fixed by
$r_{\min}=r_{0}-A$.
Hence, moderate twisting can coexist with high-frequency internal dynamics
without compromising spatial confinement, a point that will be
revisited in the discussion of wave-optics analogues (Sec .~\ref {sec:waveoptics}).

\subsection{Special Orbits and Effective Potential \label{sec:circular-orbits}}

In the above Hamiltonian discussion, one can observe three constants of motion corresponding to the cyclic coordinates $t$, $\phi$, and $z$. Thus, according to Eqs. (\ref{eq:E-const}) and (\ref{eq:L-const}), the conserved quantities ($\mathrm{E}$, $\mathrm{L}$) associated with $t$ and $\phi$ coordinates are as follows
\begin{equation}
\mathrm{E}=\dot{t}\quad,\quad \mathrm{L}=r^2 \dot{\phi} + \omega\, r p_z,
\end{equation}
where $p_z$ is given earlier. We can rewrite the above equation as
\begin{equation}
    \dot{t}=\mathrm{E},\quad \dot{\phi}=\frac{\mathrm{L}-\omega r p_z}{r^2},\quad \dot{z}=(1+\omega^2) p_z-\frac{\omega}{r} \mathrm{L}.\label{ff}
\end{equation}
Therefore, the geodesic equation for the axial coordinate $r$ using the Lagrangian (\ref{lag}) is given by
\begin{equation}
    \dot{r}^2+V_\text{eff}(r)=\mathrm{E}^2,\label{ff1}
\end{equation}
where
\begin{align}
    V_\text{eff} & =-\epsilon+\frac{\mathrm{L}^2}{r^2}+(1+\omega^2)  p^2_z-\frac{2 \omega \mathrm{L} p_z}{r}.\label{ff2}
\end{align}
which is just the effective potential given in Eq. (\ref{eq:Veff}). One can recover Minkowski flat space in the limit $\omega=0$, corresponding to the absence of the helical twist parameter. In this case, the potential is reduced to $V_\text{eff}=\left(-\epsilon+\mathrm{L}^2/r^2+p^2_z\right)$. 

Now, we specifically focus on the properties of circular photon orbits. An analogue study of circular photon orbits was carried out in Ref. \cite{EPJC2025}. To determine such orbits, the standard conditions $\dot{r} = 0 \quad \text{and} \quad \ddot{r} = 0$ must be satisfied. Applying these conditions to Eqs. (\ref{ff1}) and (\ref{ff2}), and considering the null geodesic case ($ \epsilon = 0 $), we obtain the following relations:
\begin{equation}
    \mathrm{E}^2=V_\text{eff}(r_0)=\frac{\mathrm{L}^2}{r^2_0}+(1+\omega^2) p^2_z-\frac{2 \omega \mathrm{L} p_z}{r_0},\label{ff3}
\end{equation}
with $r_0$ being the radius of the photon's circular orbit.
So, by calculating $V'_\text{eff}(r)=0$, we find
\begin{equation}
r_0=\frac{\mathrm{L}}{\omega p_z}.\label{ff4}
\end{equation}
Next, we determine $V''_\text{eff}(r)$ using the effective potential given in Eq. (\ref{ff2}) as,
\begin{equation}
    V''_\text{eff}(r)=\frac{2\mathrm{L}}{r^4}(3 \mathrm{L}-2 \omega r p_z).\label{ff5}
\end{equation}
At the radius $r=r_0$ given in Eq. (\ref{ff4}), we find
\begin{equation}
    V''_\text{eff}(r=r_0)=\frac{2 \omega^4 p^4_z}{\mathrm{L}^2}>0.\label{ff6}
\end{equation}
Moreover, the photon particle's energy at the circular null orbits using Eq. (\ref{ff3}) becomes $\mathrm{E}=p_z>0$. 

The above analysis reveals that circular null orbits are stable, and photon particles remain confined near the circular path.

\section{Deflection Angle of Photon Rays}\label{deflection}

The deflection angle is a fundamental quantity in the study of gravitational lensing, particularly in the context of cosmic strings and other exotic spacetime structures. As light propagates through curved spacetime, its trajectory is bent by the gravitational influence of intervening masses or topological defects, such as wiggly cosmic strings. This angle is a key observable, linking theoretical predictions with astrophysical observations. By analyzing its behavior, researchers can probe the geometry, mass distribution, and physical properties of lensing objects, providing a powerful means to test general relativity, detect subtle astrophysical signatures, and constrain phenomena beyond the Standard Model of cosmology \cite{vilenkin2000cosmic,dyda2007cosmic,PhysRevD.96.084047}.

In the following, we determine the deflection angle of photon rays in the gravitational field generated by the curved spacetime described by Eq.~\eqref{metric}. We employ two complementary approaches to evaluate the deflection, as detailed below:

\begin{center}
{\bf Case A: General Method}
\end{center}

The motion of light in the Riemann space-time is described by the equation $ds=0$. Using the line-element (\ref{metric}), we can write
\begin{equation}
    -\dot{t}^2 + \dot{r}^2 + (1+\omega^2)\,r^2\, \dot{\phi}^2 +2\,\omega\, r\, \dot{\phi}\,\dot{z}+ \dot{z}^2=0.\label{hh1}
\end{equation}

Let us suppose that the test particle (light) moves in the direction of the z-axis, or we can say, $r=r_0=\mbox{const}$ and $dz/dt=c=1$, where $c$ is the speed of photon particles. This is an analogy to the case studied in \cite{MP,FAA}.

Thereby, from Eq. (\ref{hh1}) we find
\begin{equation}
    (1+\omega^2) r^2_0 \left(\frac{d\phi}{dt}\right)^2 +2 \omega r_0 \left(\frac{d\phi}{dt}\right)=0.\label{hh2}
\end{equation}

Let us consider the solution of the above differential equation to be the form $\phi(t)=A\,t$, where $A$  is an arbitrary parameter that must be determined. Substituting this solution into the Eq. (\ref{hh2}) and after simplification results
\begin{equation}
    A=-\frac{2 \omega}{(1+\omega^2) r_0}.\label{hh3}
\end{equation}
Thus, the function $\phi(t)$ is in the following form:
\begin{equation}
    \phi(t)=-\frac{2 \omega}{(1+\omega^2) r_0} t.\label{hh4}
\end{equation}

Now, we choose the interval $z_2-z_1 = \Delta z =p$ (pitch); the distance between two points on the line parallel to the z-axis, then the time-interval $\Delta t$ is given by $\Delta t=p/c=p$. Thus, the deflection angle of light is given by
\begin{equation}
    \Delta \phi=-\frac{2 \omega p}{(1+\omega^2) r_0}.\label{hh5}
\end{equation}
From the expression in Eq. (\ref{hh5}), it becomes evident that the deflection of photon rays depends on the real parameter $\omega$, the distance along the $z$-axis $p$ for a fixed value of $r_0$. 

Substituting $r_0$, the deflection angle from Eq. (\ref{hh5}) becomes
\begin{equation}
    \Delta \phi=-\frac{2 \omega^2 p p_z}{(1+\omega^2) \mathrm{L}}.\label{hh6}
\end{equation}

For very small parameter $\omega <<1$, we find
\begin{equation}
    \Delta \phi \approx -\frac{2 \omega^2 p  p_z}{\mathrm{L}},\label{hh7}
\end{equation}
\noindent
which reveals that, for $\omega\ll1$, the deflection is second-order in the torsion strength and flips sign when either $\omega$ or $p_z$ changes sign.  Although suppressed by the small factor $\omega^{2}$, it can still be enhanced by large pitch $p$ or small angular momentum $L$, offering a clear geometric signature of the helical twist.

\begin{center}
    {\bf Case B: Lagrangian Method}
\end{center}

The equation of orbit is defined using Eqs. (\ref{ff}) and (\ref{ff1}) as,
\begin{equation}
    \left(\frac{dr}{d\phi}\right)^2=\frac{\dot{r}^2}{\dot{\phi}^2}=\frac{\gamma^2 r^2}{\left(\frac{\mathrm{L}}{r}-\omega p_z\right)^2}-r^2,\label{gg1}
\end{equation}
where $\gamma=\sqrt{\mathrm{E}^2-p^2_z}$. Defining a new variable via $r=\frac{1}{u}$ into the above equation results
\begin{equation}
    \frac{du}{d\phi}=\pm u \sqrt{\frac{\gamma^2}{\left(\mathrm{L} u-\omega p_z\right)^2}-1}.\label{gg2}
\end{equation}

Equation (\ref{gg2}) is the photon trajectories in the given gravitational field. One can determine the deflection angle of the photon rays using the following formula \cite{PhysRevD.97.024042}
\begin{equation}
    \hat{\alpha}=2\Delta \phi-\pi,\label{gg3}
\end{equation}
where  
\begin{equation}
    \Delta \phi=\int^{u_0}_{0}\frac{du}{u \sqrt{\frac{\gamma^2}{(\mathrm{L} u-\omega p_z)^2}-1}}.\label{gg4}
\end{equation}
Here $u_0=(\omega p_z+\gamma)/\mathrm{L}$ that is obtained from Eq. (\ref{gg2}) under the condition $(du/d\phi)|_{u=u_0}=0$.

\begin{figure*}[htbp]
  \centering
  \includegraphics[width=\textwidth]{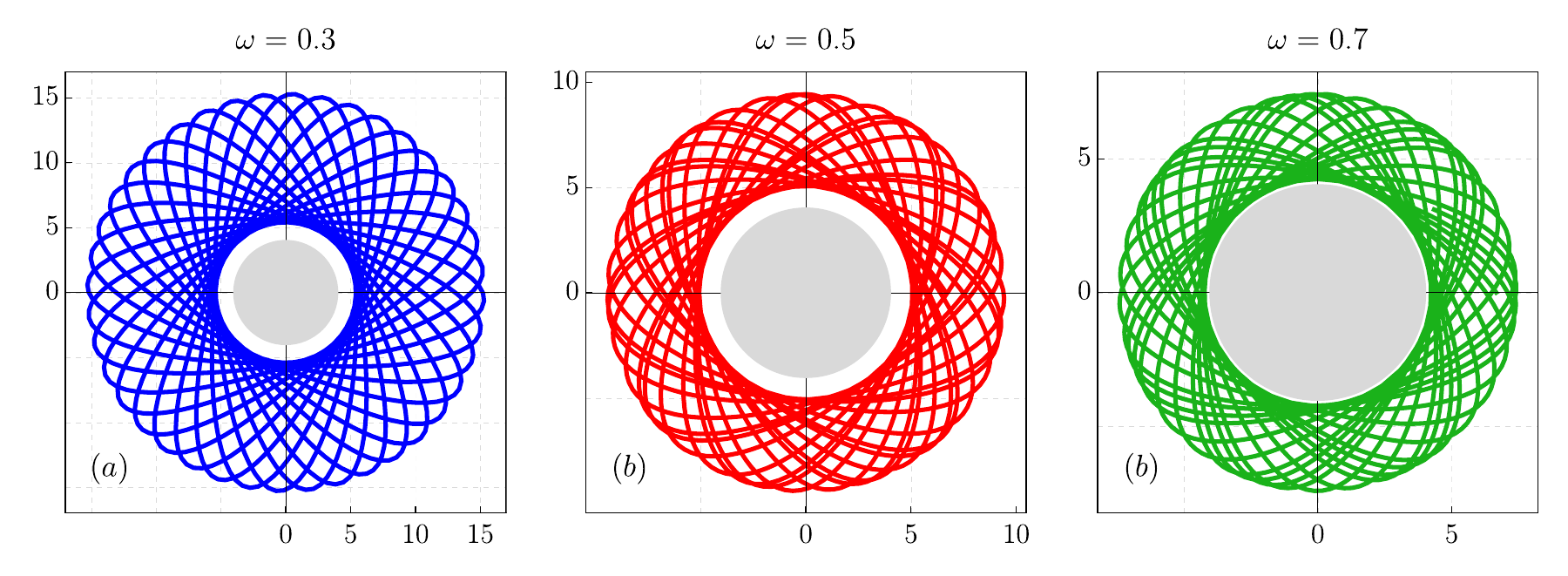}
  \caption{\footnotesize The trajectories of the 2D orbits projected onto the $x$–$y$ plane for different values of $\omega$. The inner gray circle in the third figure defines a radius of $4.05$ and represents a ``forbidden region'' where there is a barrier that the particle cannot access. In all cases we have used $\mathrm{L}=18$, $p_z=2$ and $\mathrm{E}=0.6$.}
  \label{2D}
\end{figure*}

Figure \ref{2D} shows the orbits for different values of $\omega$, obtained by taking the second derivative of Eq. \eqref{gg2} concerning $\phi$. In Fig. \ref{2D}(a), for $\omega = 0.3$, the orbit extends further outward and exhibits a larger inner free region. As $\omega$ increases, the particles move closer to the center, resulting in smaller orbital radii, see Figs. \ref{2D}(b) and (c). This behavior, in which the orbits shrink as $\omega$ increases, was anticipated in Sec. \ref{sec:circular-orbits}, where the effective geometric potential was discussed, see Eq.~\eqref{ff4}.

It is important to note that, in the trajectories obtained in this work, the presence of torsion causes the orbits to conform to Bertrand’s Theorem: they are bounded, though not necessarily closed. Additionally, the 2D orbits found here reflect the helical structure of the spacetime, which is further explored in the 3D trajectories. For 3D orbits that oscillate between a minimum and maximum radius, see Sec. \ref{sec:geodesics}.

\section{Wave optics in three-dimensional surface \label{sec:waveoptics}}

Wave optics (or physical optics) describes the behavior of light as a wave, accounting for phenomena like interference, diffraction, and polarization effects not explained by geometric (ray) optics. A central equation in wave optics is the scalar Helmholtz equation, derived from Maxwell’s equations under the assumption of monochromatic light, scalar fields \cite{PhysRevA.78.043821,PhysRevA.103.023516,PRL2010}
\begin{equation}
\nabla^2_{LB}\, \Psi(\mathbf{r}) + k^2\, \Psi(\mathbf{r}) = 0, \label{dd1}
\end{equation}
where $ \Psi(\mathbf{r}) $ is a complex scalar function (potential) defined at a spatial point $ \mathbf{r} = (x, y, z) \in \mathbb{R}^3 $, and $ k $ is a real or complex constant related to the wave properties of the medium. Here, $\nabla^2_{LB}$ explains the Laplace-Beltrami operator, and in curved space, it takes the following form:
\begin{equation}
\nabla^2_{LB}\,\Psi = \frac{1}{\sqrt{g}} \partial_i \left( \sqrt{g} g^{ij} \partial_j \Psi \right),\label{dd2}
\end{equation}
where the indices $i, j \in \{1, 2, 3\}$. Noted that we can expressed the line element (\ref{metric}) in the form: $ds^2=-dt^2+g_{ij}\,dx^i\,dx^j$, where $g_{ij}$ is the spatial part of the metric tensor $g_{\mu\nu}$ given in Eq. (\ref{metric-tensor}).

Considering the scalar function $\Psi(r, \phi, z)=\exp(i k_z z)\exp(i \ell \phi) \psi(r)$, where $k_z>0$ is the translation and $\ell$ is the angular numbers. Using this, we can re-write Eq. (\ref{dd1}) after simplification as
\begin{equation}
\psi''(r)+\frac{1}{r}\,\psi'(r)+\left[k^2_\text{eff}-\frac{\ell^2}{r^2}+\frac{2 \eta}{r}\right]\,\psi(r)=0,\label{dd3}
\end{equation}
where we defined (setting $k=\Omega\,c=\Omega$, the frequency of waves)
\begin{equation}
k_\text{eff}=\sqrt{\Omega^2-(1+\omega^2) k^2_z},\quad \eta=\omega \ell k_z.\label{dd4}
\end{equation}
Here $k_\text{eff}$ is the effective propagation constant that depends on the helical parameter $\omega$.

\subsection{Effective Potential and Refractive Index} 

Now, transforming the ansatz to a new function via $\psi(r)=R(r)/\sqrt{r}$ in Eq. (\ref{dd3}) yields
\begin{equation}
R''(r)+\left(\Omega^2-U_\text{eff}(r)\right)\,R(r)=0,\label{dd5}
\end{equation}
a Schr\"{o}dinger-like radial equation with effective potential given by
\begin{equation}
U_\text{eff}(r)=(1+\omega^2) k^2_z+\frac{\ell^2-1/4}{r^2}-\frac{2 \omega \ell k_z}{r}.\label{dd6}
\end{equation}

In the limit $\omega \to 0$, the result reduces to that of flat space. This indicates that the helical parameter $\omega$ modifies the effective potential governing the wave dynamics, distinguishing it from the flat space scenario.

\begin{figure}[htbp]
\centering
\includegraphics[width=0.9\linewidth]{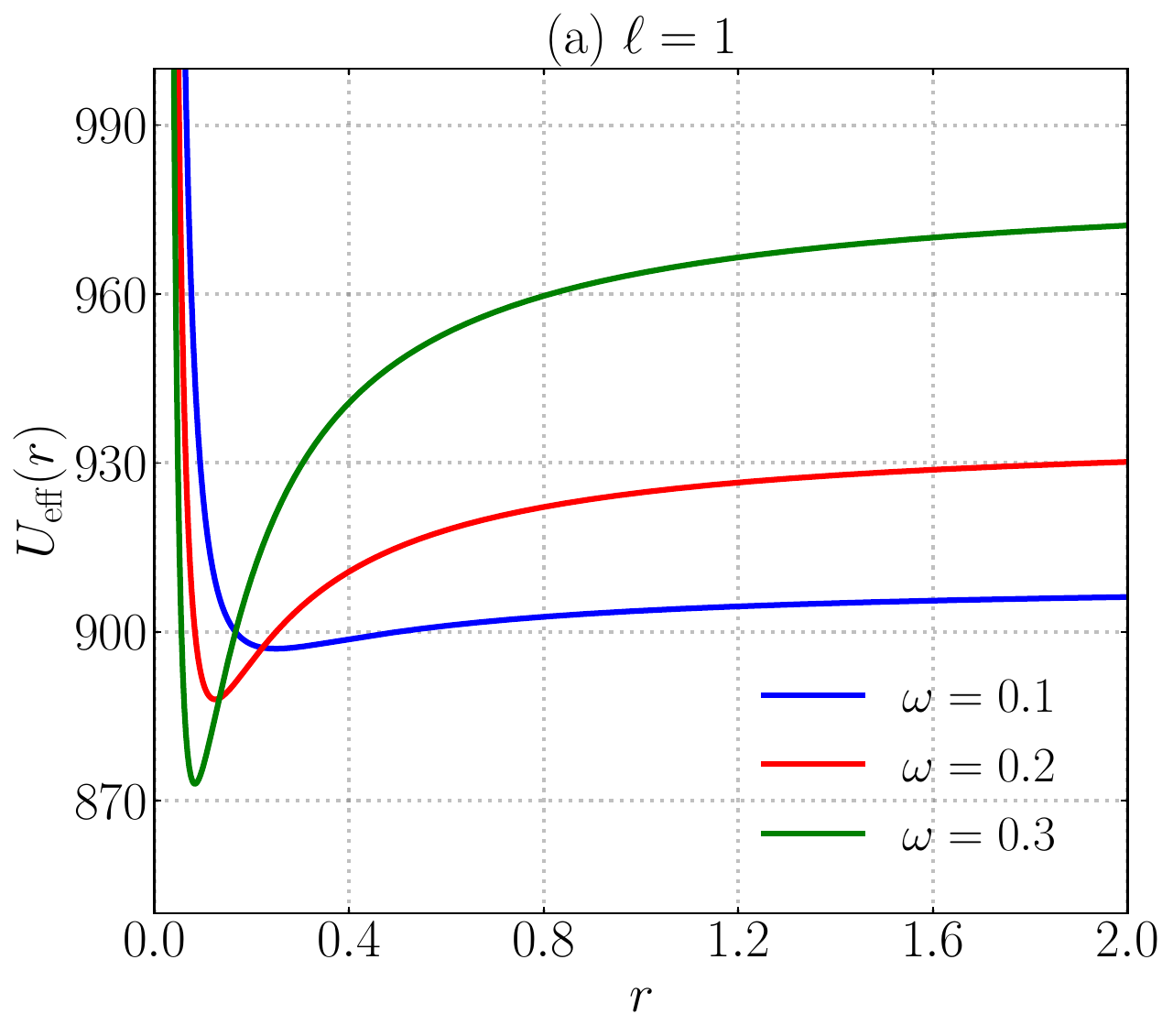}    \includegraphics[width=0.9\linewidth]{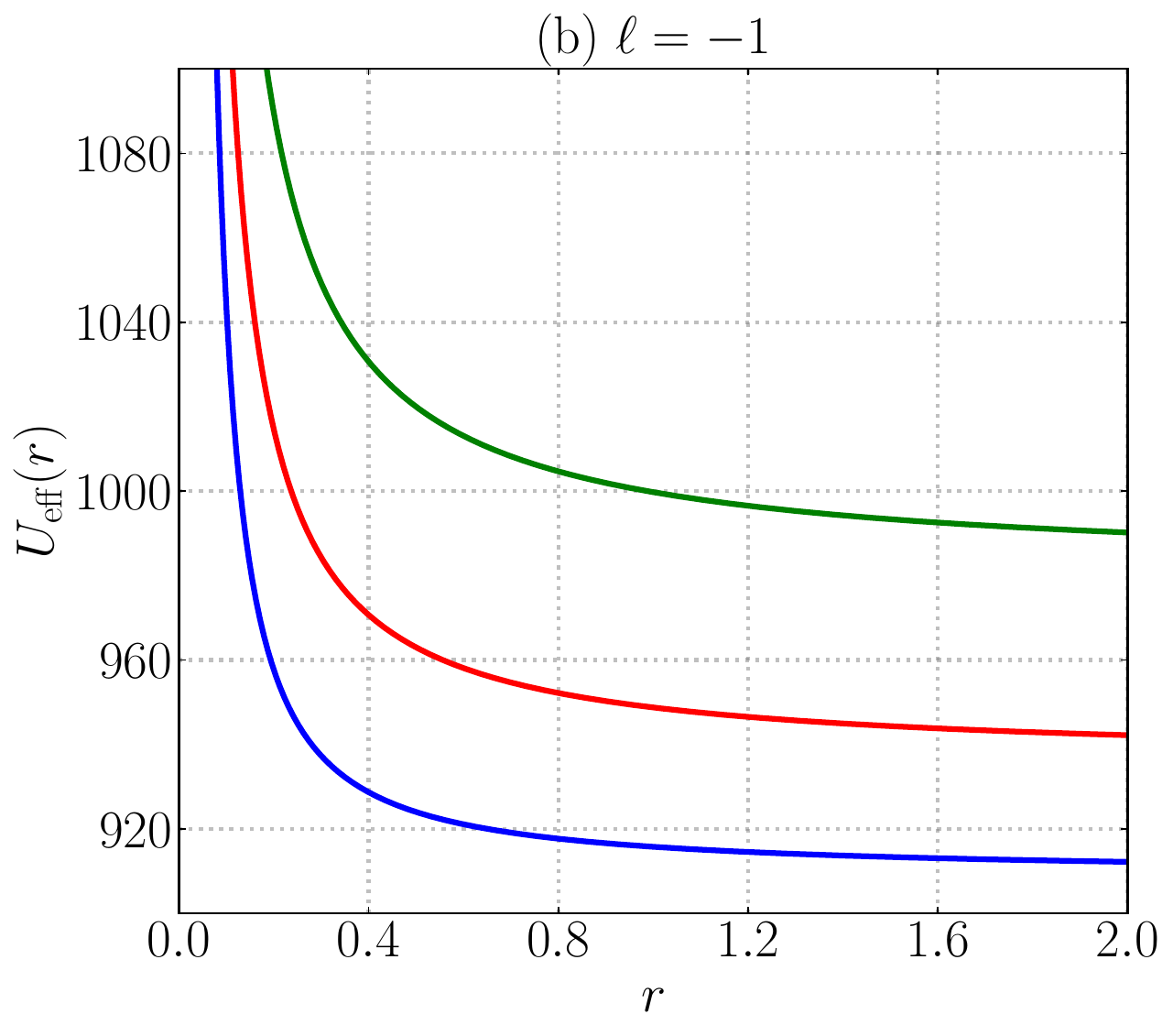}
\caption{\footnotesize Effective potential $U_{\mathrm{eff}}(r)$ given by Eq.~(\ref{dd6}) for three values of the helical parameter $\omega=0.1,0.2,0.3$ at fixed $k_z=30$. 
(a) For $\ell=1$ the term $-2\omega\ell k_z/r$ is negative, producing a short-range attractive well whose depth grows with $\omega$. 
(b) For $\ell=-1$ the same term becomes repulsive, so the potential stays positive-definite; only the centrifugal barrier $(\ell^{2}-1/4)/r^{2}$ dominates as $r\to0$.}
\label{fig:potential}
\end{figure}

\begin{figure}[htbp]
\centering
\includegraphics[width=0.97\linewidth]{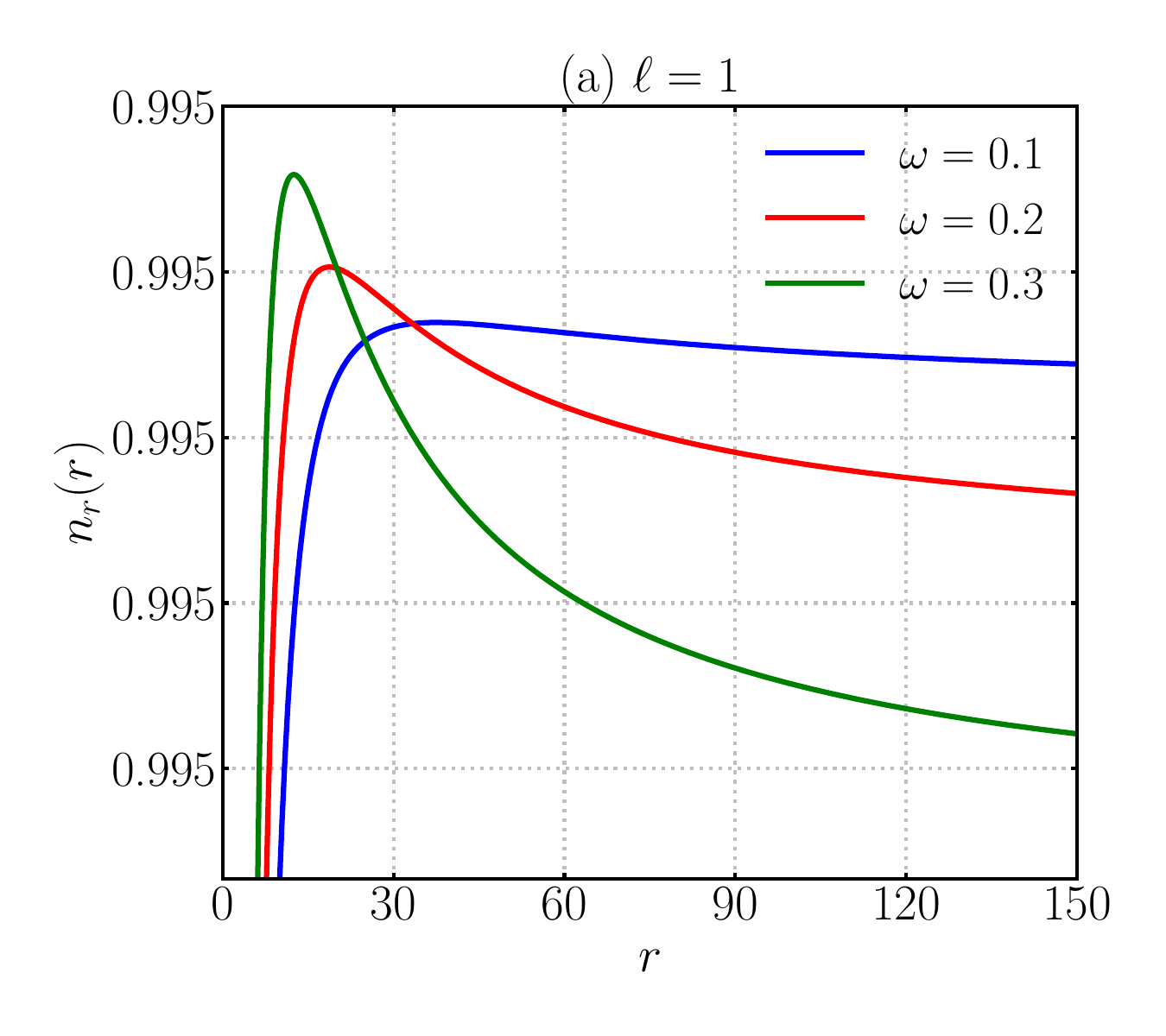}
\includegraphics[width=0.97\linewidth]{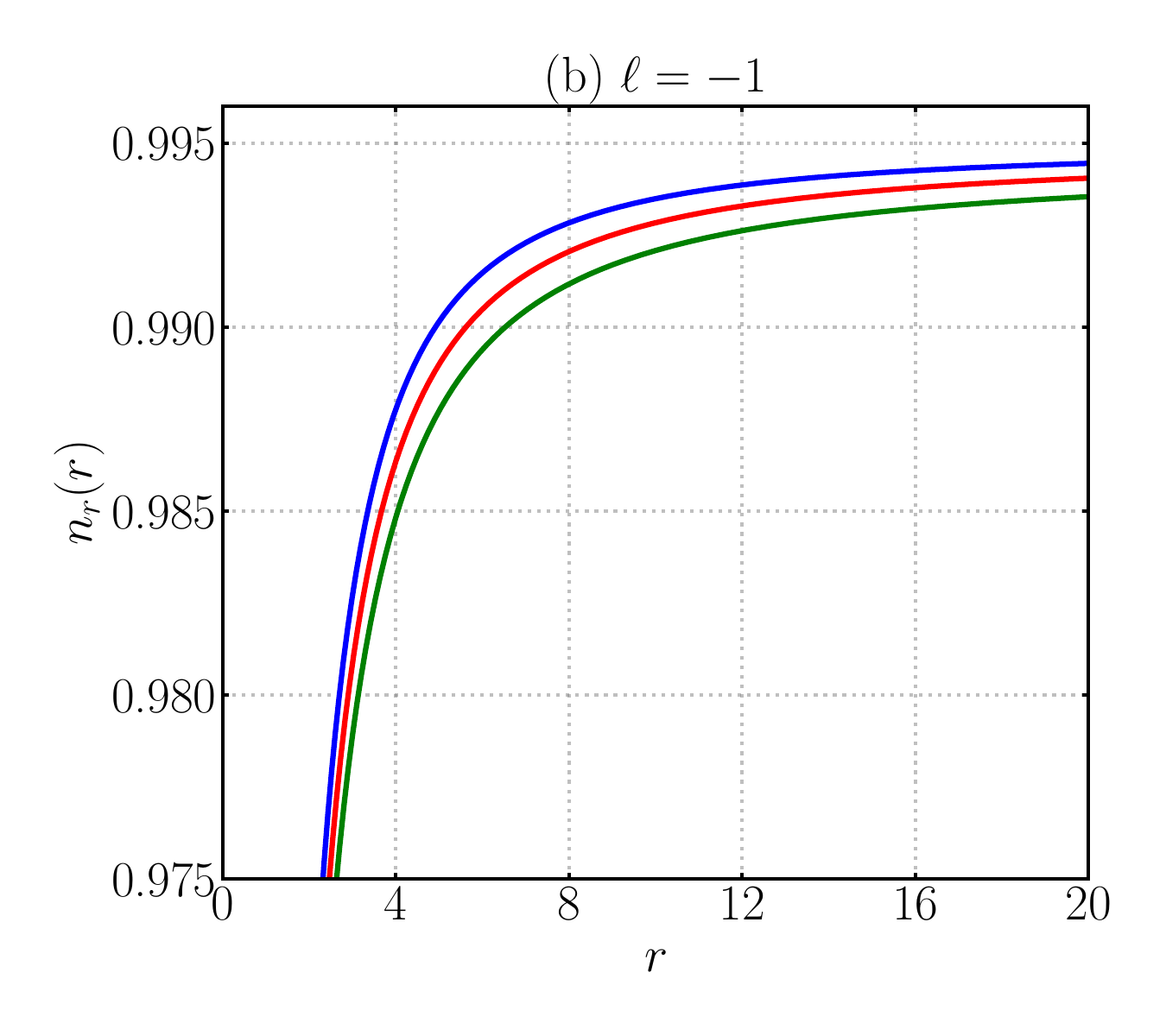}
\caption{\footnotesize Effective position-dependent refractive index $n(r)$ for
(a) $\ell = 1$ and (b) $\ell = -1$, plotted for three values of the
helical parameter $\omega = 0.1, 0.2, 0.3$.
Throughout we set $\Omega = 2$ and $k_z = 0.2$.}
\label{fig:refractive}
\end{figure}

Writing the above equation (\ref{dd5}) as 
\begin{equation}
 R''(r)+\Omega^2 n_{r}^2 R(r)=0,   
\end{equation} 
we find the refractive index of the monochromatic wave of frequency $\Omega(=k c=k)$ as
\begin{align}
n_{r}(r)&=\sqrt{1-\frac{U_\text{eff}(r)}{\Omega^2}}\nonumber\\
&=\sqrt{1-\frac{1}{\Omega^2} \left[(1+\omega^2) k^2_z+\frac{\ell^2-1/4}{r^2}-\frac{2 \omega \ell k_z}{r}\right]}.\label{dd7}
\end{align}
From the above expression, it becomes evident that several key factors influence the position-dependent effective refractive index of the electromagnetic wave. These include the helical parameter $\omega$, the angular quantum number $\ell$, and the translation quantum number $k_z$. In the limit $\omega \to 0$, the result reduces to that of flat space as
\begin{align}
n_{r}(r)=\sqrt{1-\frac{1}{\Omega^2}\,\left(k^2_z+\frac{\ell^2-1/4}{r^2}\right)}.\label{dd8}
\end{align}
When the refractive index $n_{r} > 1$, the local wave propagation is slowed down due to the influence of the curved background geometry. In such a scenario, the electromagnetic wave's phase velocity becomes less than the speed of light in vacuum ($c$), implying that the energy or information carried by a monochromatic wave or a group of waves travels more slowly than it would in free space. This condition is called the subluminal regime, where the wave propagation is impeded by the medium or effective geometry. On the other hand, when $n_{r} < 1$, the wave propagates faster than light in a vacuum. This corresponds to the superluminal regime, in which the phase velocity exceeds $c$, indicating that the wave crests move faster than they would in a vacuum. However, it is important to emphasize that even in the superluminal case, there is no causality violation, as the actual transmission of energy or information (governed by the group or signal velocity) remains constrained by the relativistic speed limit.

\begin{figure}[htbp]
\includegraphics[width=0.88\linewidth]{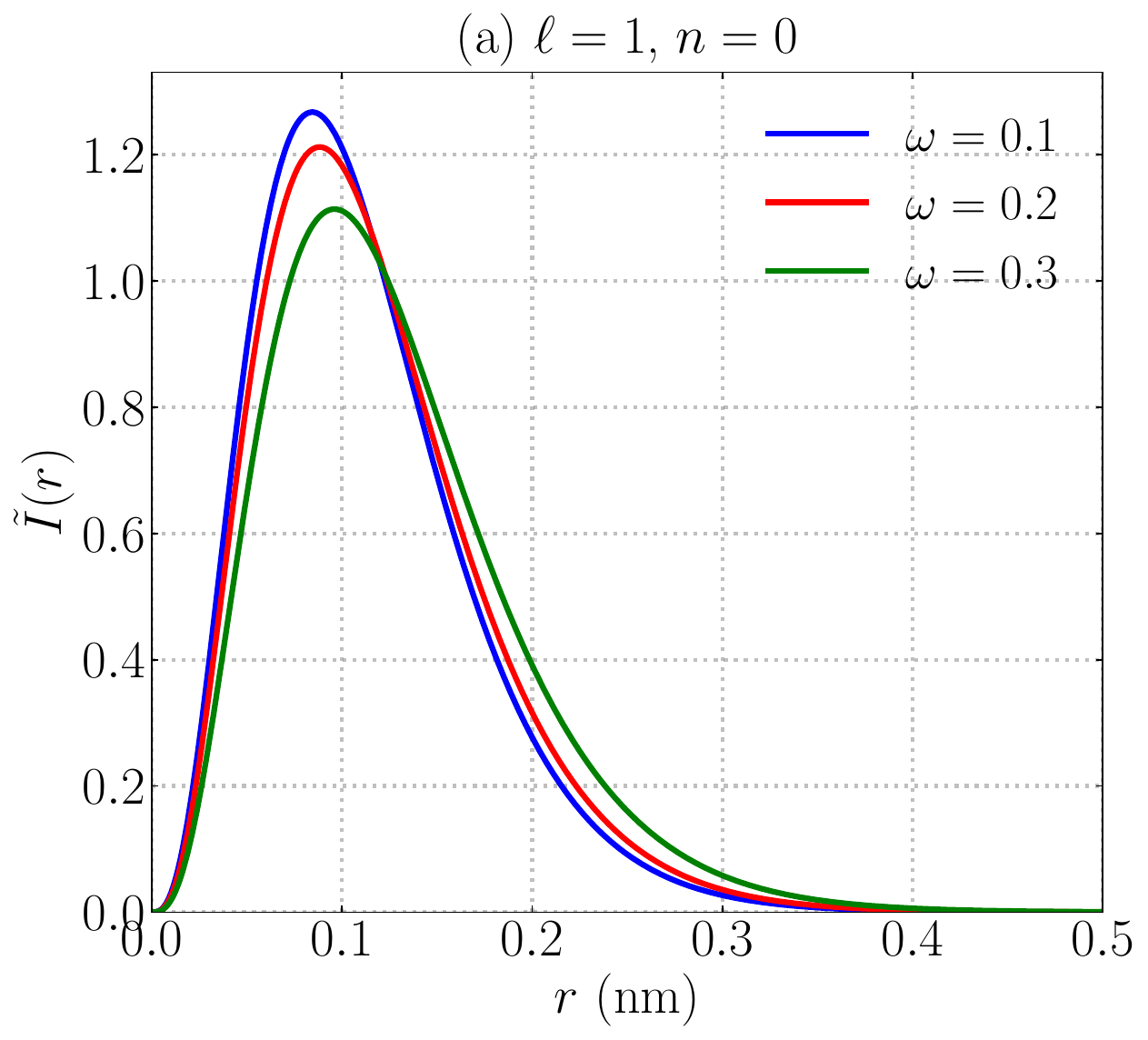}   \includegraphics[width=0.92\linewidth]{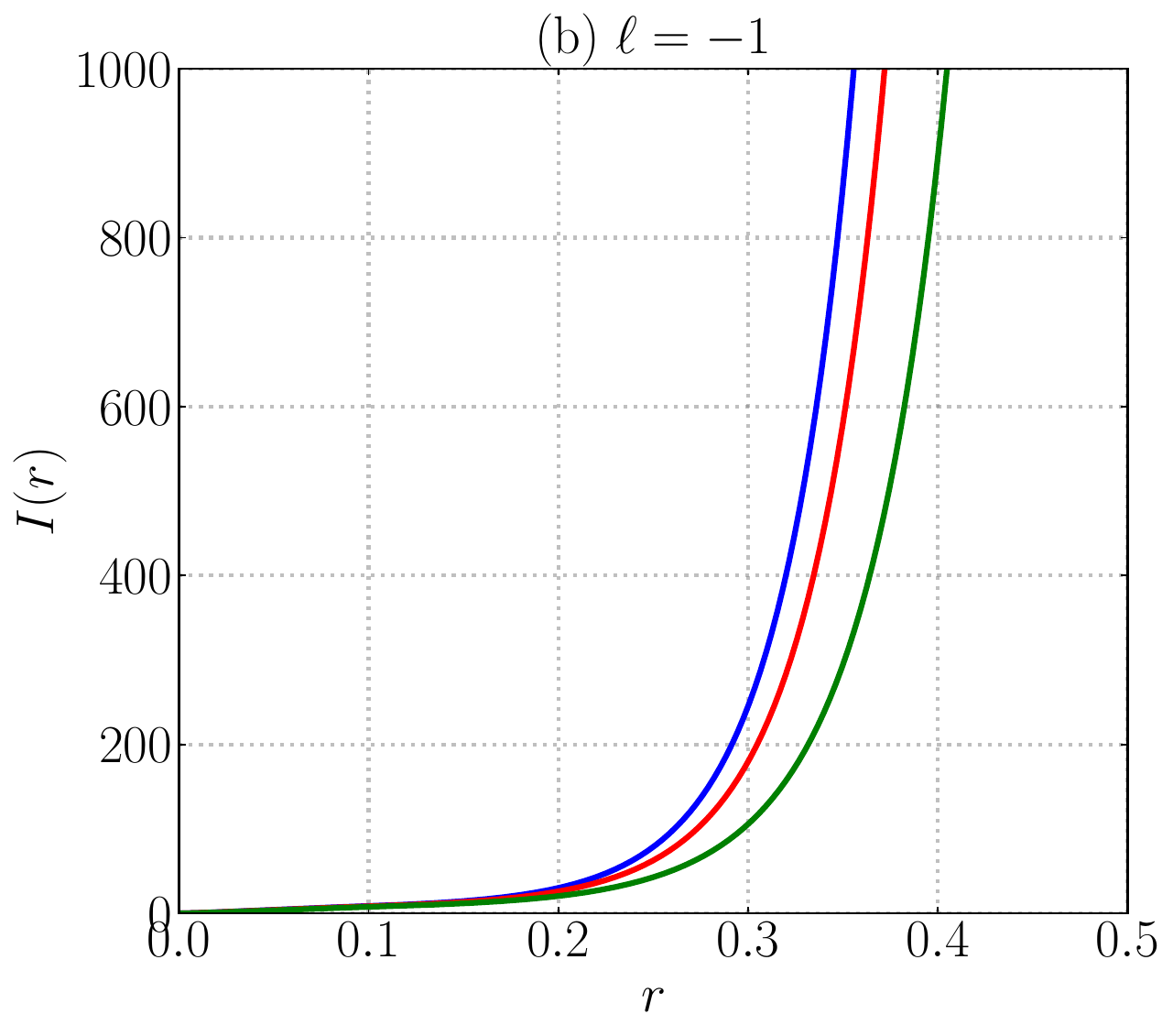}
\caption{\footnotesize
(a) Normalised radial intensity density $\tilde{I}(r)=rI(r)$ for the bound ground state ($n=0$) with azimuthal quantum number $\ell=1$ and longitudinal momentum $k_z=30$.  Three torsion strengths are shown: $\omega=0.1$ (blue), $\omega=0.2$ (red) and $\omega=0.3$ (green).
The factor $(2\lambda r)^{2\ell}$ produces the initial rise, while the Gaussian envelope $\exp[-2\lambda(\omega) r]$ enforces an exponential decay; the peak therefore shifts only slightly as $\omega$ varies. (b) Intensity density for the scattering solution with azimuthal number $\ell=-1$ and the same $k_z$ and $n$.  Here, the hypergeometric parameter remains at its generic value $a=1/2$, so the mode is not square-integrable: $I(r)$ decreases near the axis but eventually grows exponentially, signalling an outgoing wave.  Together, the two panels illustrate how changing the sign of $\ell$ turns the torsion-coupling term $-\,2\omega\ell k_z/r$ from attractive (panel~(a)) to repulsive (panel~(b)), thereby converting a confined Gaussian-like mode into a delocalised scattering state while leaving the background geometry unchanged.}
\label{fig:intensity_bound_scatter}
\end{figure}

\begin{figure*}[htbp]
\includegraphics[scale=0.34]{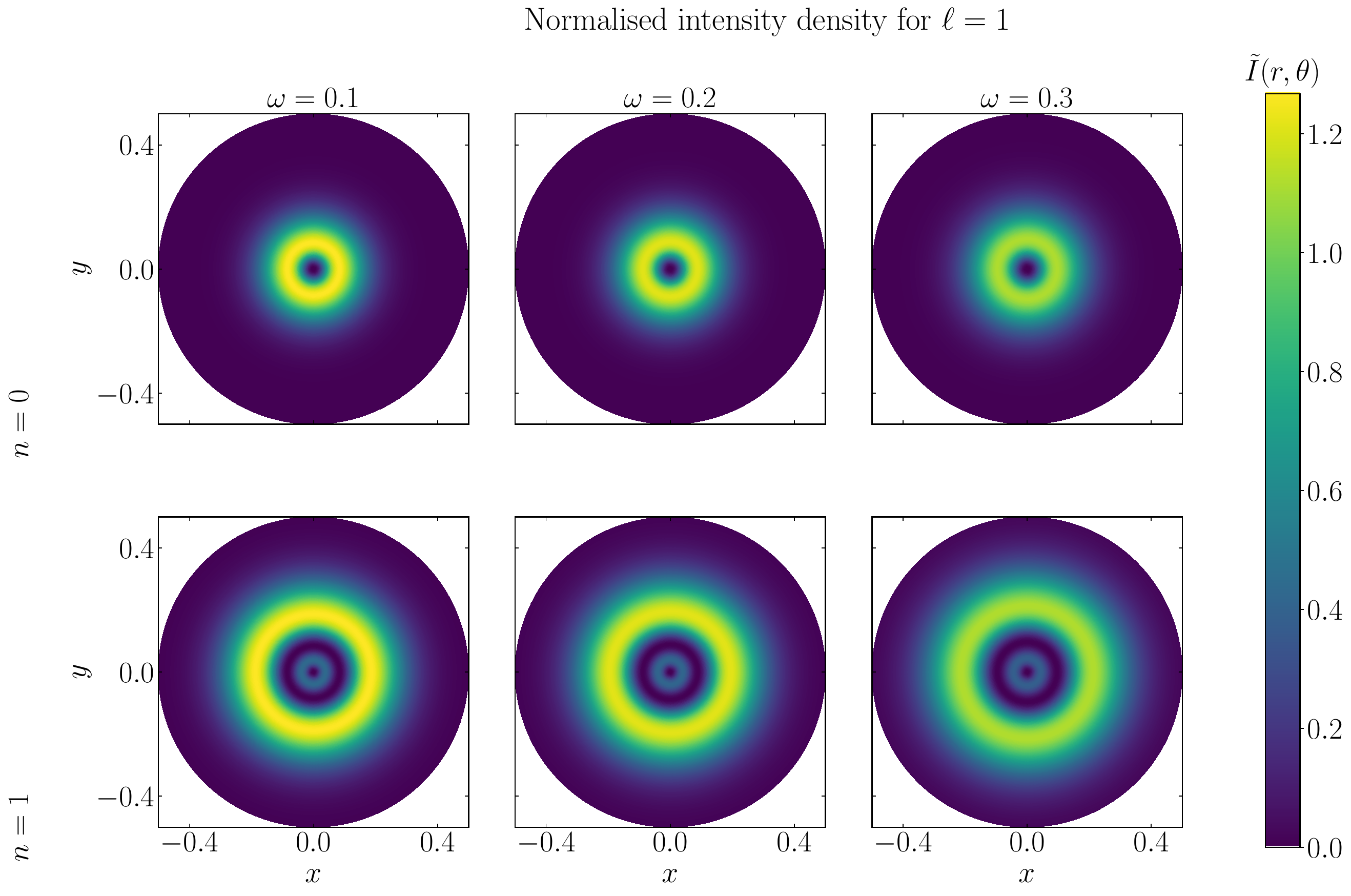}
\caption{\footnotesize
Normalised intensity density $\tilde{I}(r)$ in the transverse $(x,y)$ plane for azimuthal quantum number $\ell = 1$. Each panel is a polar pseudocolour map from the full analytical solution~\eqref{dd13}. Columns: three torsion strengths
$\omega = 0.1,\;0.2,\;0.3$ (left to right). Rows: bound ground state $n=0$ (top) and first excited state $n=1$ (bottom).
For each row, the colour scale is normalised independently so that the brightest yellow corresponds to the peak of that $n$-manifold; the vertical bar on the right therefore shows the actual scale for the top row ($n=0$). For $n=0$ the mode is single-lobed and Gaussian-like, while for $n=1$ a clear nodal ring appears at $r\simeq0.17$, as expected from the factor $(2\lambda r)^{2\ell}$ and the first zero of the confluent hypergeometric function. Increasing $\omega$ raises the propagation constant $\lambda(\omega)$, shrinking the transverse width of both states without changing their qualitative structure.}
\label{fig:intensity_2d_grid}
\end{figure*}

In Figure~\ref{fig:potential}, we present a few plots illustrating the behaviour of the effective potential $U_{\text{eff}}(r)$ in the context of wave optics for various values of the helical parameter $\omega$. The top panel corresponds to the case $\ell = 0$, while the bottom panel shows the results for $ \ell = 1$. In the top panel, the effective potential is observed to increase gradually with increasing $r$, and this overall trend shifts upward as the value of $\omega$ increases. In contrast, in the bottom panel, the effective potential decreases monotonically with $r$, and this behaviour shifts downward for larger values of $ \omega $. These contrasting trends highlight the significant role played by both the angular momentum parameter $ \ell $ and the helical parameter $ \omega $ in shaping the effective potential of the wave dynamics.

In Fig.~\ref{fig:refractive} we display the radial profile of the effective refractive index $n_{r}(r)$ at fixed monochromatic frequency $\Omega = 2$ and translational quantum number $k_z = 0.2$.
Fig. (a) corresponds to the angular-momentum sector $\ell = 1$, while Fig. (b) shows the result for $\ell = -1$. In both cases, the curves lie in the superluminal regime ($n_r < 1$), with the curves approaching a nearly constant value. In Fig. (a), for small $\omega$, this asymptotic value is close to unity as $r$ increases. In contrast, as $\omega$ increases, the curves approach lower values and may even intersect those with smaller torsion values.
The slight differences between the two panels stem from the sign of $\ell$: for $\ell = 1$ the centrifugal term raises the effective index near the core, whereas for $\ell = -1$ the combined action of the helical coupling and the negative angular momentum lowers it, producing a deeper minimum.
Reducing $k_z$ below $\sim0.14$ would invert this behaviour, driving the system into the subluminal regime ($n_{r}>1$) over a finite radial interval.

\subsection{Solution and Intensity Profiles}

To investigate the solution of Eq. (\ref{dd3}), we will consider a transformation of the radial coordinate and the function $\psi(r)$ as,
\begin{equation}
\rho = 2k_\text{eff}\, r, \quad \psi(r) = \rho^{|\ell|}\, e^{-\rho/2}\,f(\rho).\label{dd9}
\end{equation}
Substituting this into the equation (\ref{dd3}) leads to the standard confluent hypergeometric differential equation for $f(\rho)$ as,
\begin{equation}
\rho\,f''(\rho)+\left(\mathrm{b} - \rho \right)\,f'(\rho)-\mathrm{a}\,f = 0,\label{dd10}
\end{equation}
whose solution is given in terms of the confluent hypergeometric function
\begin{equation}
\psi(r) = (2k_\text{eff} \,r)^{|\ell|}\,e^{-k_\text{eff} r}{}_1 F_1\left(\mathrm{a},\mathrm{b},\, 2k_\text{eff} \,r\right).\label{dd11}    
\end{equation}

Therefore, the complete scalar complex function $\Psi$ is given by
\begin{align}
\Psi(r,\phi,z)=e^{ik_zz}\,e^{i\ell\,\phi}\,e^{-kr}(2k_\text{eff}\,r)^{|\ell|}{}_1 F_1\left(\mathrm{a},\mathrm{b},2k_\text{eff}\,r\right),\label{dd12} 
\end{align}
where ${}_1 F_1 (\mathrm{a},\,\mathrm{b},\,\rho)$ is the confluent hypergeometric function,  and 
\begin{equation}
\mathrm{a}=|\ell| + \frac{1}{2} - \frac{\omega\,\ell\,k_z}{\sqrt{\Omega^2-(1+\omega^2)\,k^2_z}},\quad \mathrm{b}=1+2\,|\ell|.\label{dd14}
\end{equation}
The intensity of the beam of waves propagating in the curved space is calculated using the relation
\begin{align}
I(r)=\Psi^{*}\,\Psi
=4\,e^{-2k_\text{eff}r}\,(k_\text{eff}r)^{2 |\ell|}\left[{}_1 F_1(\mathrm{a},\,\mathrm{b},2k_\text{eff}\,r)\right]^2.\label{dd13}
\end{align}
In the limit $\omega=0$, which corresponds to the Minkowski flat space, the beam intensity reduces as (setting $k_z=0$ due to the cylindrical symmetry)
\begin{align}
I(r)=4\,e^{-2 \Omega r} (\Omega r)^{2 |\ell|} \left[{}_1 F_1\left(|\ell| + \frac{1}{2},1+2 |\ell|, 2 \Omega r\right)\right]^2.\label{dd15}
\end{align}

From the expression (\ref{dd13}), it becomes evident that several key factors influence the beam intensity propagating in a curved space background. These include the helical torsion parameter $\omega$, the angular quantum number $\ell$, and the longitudinal momentum $k_z>0$. Notably, due to the coupling between $\omega$ and $k_z$, the case $k_z = 0$ is not physically meaningful and therefore cannot be considered. Additionally, the wave frequency $\Omega$, which corresponds to the propagation constant $k = 2\pi/\lambda$ (where $\lambda$ is the wavelength), also significantly determines the beam intensity. The intensity undergoes a modification due to the helical parameter $\omega$, compared to the flat space result presented in (\ref{dd15}).  

Figure \ref{fig:intensity_bound_scatter}(a) confirms that, once the quantisation condition $a=-n$ is imposed, the optical mode acquires a genuinely Gaussian–type envelope even in the presence of torsion.  
For $\ell=1$, the peak of the shell-normalised radial intensity density, $\tilde{I}(r)$, occurs near $r\approx0.05$ and decays on the scale $\lambda^{-1}$, matching the depth and position of the potential minima in Fig.~\ref{fig:potential}(a).  
The decrease in peak height with $\omega$ shows that torsion redistributes intensity within essentially the same radial window, preserving transverse confinement.  
The apparent ``centrifugal'' effect is, in fact, a subtle consequence of enhanced centripetal confinement due to the torsion parameter $\omega$. As seen before, the effective potential deepens with increasing $\omega$ for $\ell = 1$, reinforcing localization near the axis. However, the peak of the ground-state intensity remains nearly unchanged because the sharper potential is balanced by faster exponential decay of the wavefunction, governed by $\lambda(\omega)=\sqrt{\Omega^{2}-(1+\omega^{2})k_{z}^{2}}$. This interplay highlights how torsion reshapes mode confinement without significantly shifting the peak, thereby blending classical and wave-optical effects in a nontrivial manner.
In contrast, Fig.~\ref{fig:intensity_bound_scatter}(b) corresponds to $\ell=-1$, for which the coupling term is repulsive and the effective potential of Fig.~\ref {fig:potential}(b) remains strictly positive.  
With no well to support bound states, the wave function retains its generic parameter $ a=\tfrac {1} {2} $ and becomes non-normalizable: $I(r)$ decays close to the axis but diverges exponentially once the centrifugal barrier is surpassed, a clear fingerprint of a scattering solution.  
The side-by-side comparison emphasises that changing the sign of the azimuthal quantum number toggles the torsion term from attractive to repulsive, transforming a tightly confined beam into a delocalised outgoing mode without altering the underlying metric.

Figure~\ref{fig:intensity_2d_grid} provides a two-dimensional visualisation of the shell-normalised density $\tilde{I}(r)$ discussed above. The polar maps highlight two features that the one-dimensional profiles of Fig.~\ref{fig:intensity_bound_scatter} suggest but cannot display directly. First, the ground state ($n=0$, top row) remains single-lobed and Gaussian-like even in the presence of torsion, showing that the factor $(2\lambda r)^{2\ell}$ does not introduce additional nodes when $\ell = 1$. Second, the first excited state ($n=1$, bottom row) exhibits a clear nodal ring whose radius is fixed almost entirely by the first zero of the confluent hypergeometric function ${}_{1}F_{1}$; torsion merely rescales the
envelope through the propagation constant $\lambda(\omega)$. For visual clarity, the colour scale is normalised independently in each row, so the brightest yellow corresponds to the peak value of that $n$-manifold; absolute heights therefore match the radial plots of Fig.~\ref{fig:intensity_bound_scatter}.
The gentle radial contraction from left to right quantifies the effect of increasing $\omega$: a larger torsion parameter raises $\lambda$, tightening the transverse confinement while leaving the angular structure untouched. Combined with Figs.~\ref{fig:potential} and \ref{fig:intensity_bound_scatter}, this two-dimensional portrait completes the correspondence between the geometric potential, the radial density, and the full spatial distribution of the confined optical modes.

\section{Concluding Remarks}
\label{sec:conclusion}

We have introduced and analysed an exact, stationary, cylindrically symmetric spacetime whose geometry incorporates a built-in helical twist through the parameter $\omega$.  The metric \eqref{metric} leads to an analytic energy–momentum tensor that is negative near the axis, violates the weak energy condition, and remains anisotropic everywhere.  Because every component decays at least as $r^{-2}$, the exotic matter required to sustain the twist becomes negligible at large distances, so ordinary matter dominates in the asymptotic region.

The dynamics of test particles are governed by the effective potential \eqref{eq:Veff}, where the coupling term $-\omega \mathrm{L}p_{z}/r$ competes with the usual centrifugal barrier.  Depending on the sign of $\omega p_{z}$, the background can either confine or repel particles, yielding stable circular photon orbits and genuine helical trajectories for massive probes. A Hamiltonian treatment supplemented by Poincaré maps reveals that increasing $|\omega|$ drives the system smoothly from regular motion to a mixed, quasi-periodic regime, thereby establishing $\omega$ as an effective control parameter for orbital architecture.

In the wave optics sector, the background acts as a radially varying medium whose refractive index~\eqref{dd7} depends on $\omega$, on the angular number $\ell$, and on the longitudinal momentum $k_{z}$.  Even moderate torsion strengths ($|\omega|\lesssim 0.3$ in our normalisation) produce sizeable shifts in the propagation constant and marked intensity modulations, which suggests that photonic or phononic analogues of the metric could be engineered in metamaterials.  The same coupling also affects light deflection: for small $\omega$, the bending angle grows linearly with the pitch and quadratically with $\omega$, providing a clear observable that distinguishes this spacetime from straight cosmic strings or torsion-free helicoids.

Several avenues remain open.  Coupling spinor or gauge fields to the background might clarify whether the exotic source can be interpreted as an effective fluid endowed with internal degrees of freedom.  Allowing $\omega$ to vary in time or space would turn the solution into a toy model for rotating cosmic defects and could reveal parametric instabilities in the geodesic flow.  Finally, quantising matter and gravitational perturbations on this geometry may uncover torsion-driven vacuum-polarisation or Casimir effects with potential laboratory analogues.  Taken together, our results show that helically twisted spacetimes provide a fertile arena to explore the interplay between curvature, torsion, and matter, both in fundamental gravity and condensed-matter analogue systems.
}

\section*{Acknowledgments}
{\footnotesize This work was supported by CAPES (Finance Code 001), CNPq (Grant 306308/2022-3), and FAPEMA (Grants UNIVERSAL-06395/22 and APP-12256/22). F. A. acknowledges the Inter University Center for Astronomy and Astrophysics (IUCAA), Pune, India, for granting a visiting associateship.
}

\bibliographystyle{apsrev4-2}
%

\end{document}